\documentclass[aps,prx,twocolumn,superscriptaddress,longbibliography]{revtex4-2}

\usepackage{graphicx,amsmath,amssymb}
\usepackage{float}

\newcommand{\mysec}[1]{\section*{#1}}
\newcommand{\mysubsec}[1]{\subsection{#1}}

\makeatletter
\renewcommand{\p@subsection}{}
\makeatother

\newcommand{\change}[1]{{#1}}

\usepackage{hyperref}
\usepackage[nameinlink]{cleveref}
\Crefname{equation}{Eq.}{Eqs.}
\Crefname{figure}{Fig.}{Figs.}
\Crefname{efigure}{Extended Data Fig.}{Extended Data Figs.}
\Crefname{tabular}{Tab.}{Tabs.}
\Crefname{linenum}{line}{lines}
\crefrangelabelformat{equation}{#3#1#4-#5#2#6}
\creflabelformat{equation}{#2#1#3}

\Crefformat{section}{Methods #2#1#3}

\begin{document}


\title{Dynamical stability for dense patterns in attractor neural networks}


\author{Uri Cohen}
\email{corresponding author: uc231@cam.ac.uk}
\affiliation{Computational and Biological Learning Lab, Department of Engineering, \\University of Cambridge, Cambridge, UK}

\author{Máté Lengyel}
\affiliation{Computational and Biological Learning Lab, Department of Engineering, \\University of Cambridge, Cambridge, UK}
\affiliation{Center for Cognitive Computation, Department of Cognitive Science, \\Central European University, Budapest, Hungary}




\begin{abstract}
%
Recurrent neural networks are canonical models of biological memory. In these models, memories are represented by distributed patterns of neural activity that are stored in the recurrent connections between neurons, such that they become attractors of the network's dynamics. During memory recall, network dynamics thus converge toward one of these memory patterns when started from a noisy or partial cue. Therefore, memory performance critically hinges on the dynamical stability of the stored patterns. However, previous theoretical approaches only studied dynamical stability under highly restrictive conditions that do not readily apply to biological neural circuits. Here, we develop a theory of the local stability of discrete fixed points in a broad class of networks with graded neural activities and in the presence of noise. Using methods from random matrix theory, we analyze the bulk and outliers of the eigenvalue spectra of the Jacobians that characterize network dynamics around fixed points. We show that either all fixed points are stable or all of them are unstable, depending on whether their number is below a ``critical load for stability'', which is distinct from the classical critical capacity that measures the maximal number of achievable fixed points regardless of their stability. We further analyze the dependence of this critical load for stability on experimentally measurable quantities characterizing the statistics of memory patterns and the activation functions of neurons. Our analysis highlights the computational benefits of sparse-like patterns and threshold-linear activation functions and offers testable predictions for neural circuits supporting memory.

\vspace{3pt}
Significance statement: How our brains store and retrieve memories is poorly understood. Previous theoretical approaches made assumptions about the properties of neural circuits that were incompatible with biology.
We develop a theory, based on random matrix theory, to analyze the capacity of neural networks for reliably recalling previously stored memories under minimal, \change{more} biologically plausible assumptions about neural circuit properties. 
We find that neural networks can reliably recall either all or none of the memories they store. The theory makes non-trivial, experimentally testable predictions about the properties of memory patterns and neural activation functions.  
Our work \change{helps reveal}  the principles underlying how networks of interacting neurons \change{might} perform one of the most impressive feats of the brain: the remembrance of things past.

\vspace{3pt}
\noindent auto-associative memory $|$ attractor networks $|$ dynamical stability $|$ memory recall $|$ random matrices

\end{abstract}

\maketitle



\noindent Attractor neural networks are canonical models of biological memory: they store neural activity patterns as stable fixed points (i.e., attractors) in their connection weights, so that when started from a noisy or incomplete version of one of these memory patterns as an initial condition, their autonomous dynamics converge to the corresponding fixed point owing to its dynamical stability -- thus performing `auto-associative' memory recall \cite{hopfield1982neural, amit1985spin}. 
Therefore, the stability of memory patterns as fixed points is critical for the operation of attractor networks. However, previous approaches had limited success in studying fixed-point stability. 
The `Hebbian' approach guarantees fixed-point stability in attractor networks by constructing an energy (or Lyapunov) function that is minimized by the network dynamics \cite{hopfield1982neural}. This has only been possible in a few (albeit very successful) cases \cite{kanter1987associative,tsodyks1988enhanced,treves1990graded,lengyel2005matching,krotov2016dense} after making specific assumptions about the statistics of memory patterns (typically assumed to be binary), single-neuron activation functions (saturating, or rectified-linear), and in particular the way memory patterns influence dynamics (through some form of a so-called `Hebbian' learning rule) requiring normal connection weight matrices. Even when those assumptions were violated, stability was achieved by approximately following such an energy function \cite{kree1987continuous,treves1988metastable,tirozzi1991chaos}.
Conversely, the `Gardner' approach allows the analysis of the storage capacity of neural networks in terms of the number of fixed points that can be embedded in their dynamics without recourse to an energy function \cite{gardner1988space, schonsberg2021efficiency}, but remains entirely mute about the stability of the embedded fixed points. Finally, optimization-based numerical approaches have also been used to embed stable fixed points in neural networks without making limiting assumptions, but they did not lend themselves to theoretical insight \cite{festa2014analog}.

Here, we extend the `Gardner' approach to gain analytical insights about the stability of fixed points in a broad class of networks with graded neural activities and generic, non-saturating, rectified, power-law activation functions.
In particular, rather than the oft-studied sparse limit, here we consider dense patterns for the analytical tractability of dynamical stability, and also because the inherent noisiness of neural signaling can easily prevent firing rates from being exactly zero in practice. To supplement these analyses, we also consider sparse patterns 
and show that our results extend to those in numerical simulations.
We demonstrate that there is a phase transition for stability in such networks: optimizing network connectivity to maintain memory patterns as fixed points with minimal weights renders either all or none of those fixed points stable, depending on pattern statistics and single-neuron properties. 
By characterizing the conditions under which fixed-point stability emerges in a large class of network dynamics, we provide design principles for biological systems performing auto-associative memory \cite{treves1992computational, treves1996much}.

\begin{figure*}
\centering
\includegraphics[width=\linewidth]{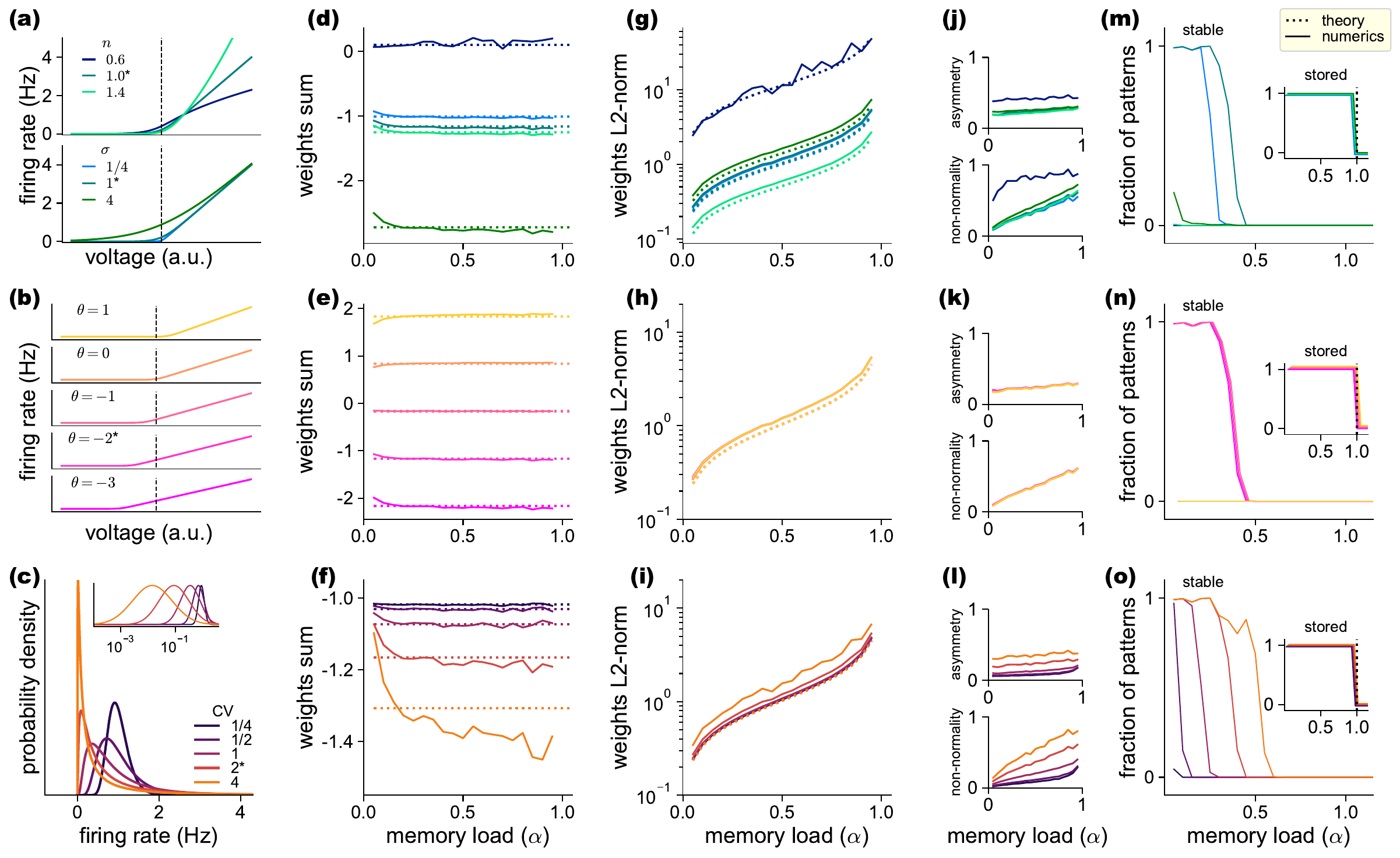}
\caption{\label{fig:capacity}
\textbf{Storing fixed-points.} \change{
(a) The activation function for different choices of the exponent, $n$ (top), and smoothness, $\sigma$ (bottom, color coded). 
(b) The activation function for different choices of the threshold, $\theta$ (color coded). 
(c) Probability density of log-normal distributed patterns at different values of CV (color coded). The inset shows normalized density on a log firing-rate scale. 
(d) The sum of the presynaptic weights of a neuron (i.e.\ a row of $\mathbf{W}$) as a function of memory load ($\alpha$, x-axis) for different activation functions (colors as in (a)). 
(e-f) Same as (d), using parameter combinations and color code as (b) and (c), respectively.
(g) L2-norm of the presynaptic weights of a neuron (i.e.\ a row of $\mathbf{W}$) as a function of memory load ($\alpha$, x-axis) for different activation functions (colors as in (a)). Note the divergence of the L2-norm as $\alpha\rightarrow1$, as predicted by theory. 
(h-i) Same as (g), using parameter combinations and color code as (b) and (c), respectively.
(j) Measures of the weights asymmetry (top) and dynamics non-normality (bottom) as a function of memory load ($\alpha$, x-axis), for different activation functions (colors as in (a)).
(k-l) Same as (j), using parameter combinations and color code as (b) and (c), respectively.
(m) The fraction of patterns which are correctly stored (inset) or stable for recall as a function of memory load ($\alpha$, x-axis) for different activation functions (colors as in (a)).
(n-o) Same as (m), using parameter combinations and color code as (b) and (c), respectively. 
Solid vs.\ dotted lines in (d-l) and the insets of (m-o) show numerical vs.\ theoretical results (with only numerical results shown for stability in (m-o) -- see subsequent figures for corresponding theoretical results). For each combination of parameters, weights $\mathbf{W}$ of networks of size $N=512$ were numerically optimized according to \Cref{eq:optW}, with $P=\left\lfloor\alpha\,N\right\rfloor$.
Starred parameter values in (a-c) indicate default settings, used in all simulations but those for which the results are reported with the given parameter being color coded.}
}
\end{figure*}

\mysec{Network model for auto-associative memory}

\newcommand{\ratefunv}{\mathbf{h}}
\newcommand{\ratefun}{h}

We study a network of $N$ neurons with voltage dynamics:\looseness=-1
\begin{equation}\label{eq:dynamics}
    \tau\,\dot{\mathbf{v}} = -\mathbf{v}+\mathbf{W}\,\ratefunv\!\left(\mathbf{v}\right)
\end{equation}
where $v_i$ is the voltage of neuron $i$, $\tau$ is the neural time constant, $\mathbf{W}\in\mathbb{R}^{N\times N}$ defines recurrent connection weights (such that $W_{ij}$ is the strength of the connection from neuron $j$ to neuron $i$), and $\ratefun_i\!\left(\mathbf{v}\right)=\ratefun\!\left(v_i\right):\mathbb{R}\rightarrow\mathbb{R}^+$ is the neural activation function that maps the `voltage' of a neuron, $v_i$, to its (positive) instantaneous firing rate. 
(Note that $\tau$ and $\ratefun\!\left(\cdot\right)$ are shared across neurons.)
Defining $\mathbf{v}=\mathbf{W}\,\mathbf{r}$, \Cref{eq:dynamics} has an equivalent form of rate dynamics \cite{miller2012mathematical}, which is the form we will use in the following for mathematical convenience: 
\begin{equation}\label{eq:dynamics_r}
\tau\,\dot{\mathbf{r}} = -\mathbf{r} + \ratefunv\!\left(\mathbf{W} \,\mathbf{r}\right) 
\end{equation}

While our theoretical results hold for a wide range of activation functions, $\ratefun\!\left(\cdot\right)$, with additional assumptions detailed below, we consider here the specific case of the soft-rectified power-law to be able to systematically study how stability depends on its parameters:
\begin{equation}\label{eq:noisy_nonlinearity}
    \ratefun\!\left(v\right) = g\!\left(v-\theta\right) \text{, where } g\!\left(v\right) = \left[\frac{\sigma}{\pi} \, \ln\!\left(1+e^{\frac{\pi}{\sigma}\,v}\right)\right]^n 
\end{equation}
\change{with exponent $n$ (\Cref{fig:capacity}a, top), smoothness $\sigma$ (\Cref{fig:capacity}a, bottom), and threshold (or negative bias) $\theta$ (\Cref{fig:capacity}b)}. 
The particular form of \Cref{eq:noisy_nonlinearity} is motivated as an approximation to a hard-rectified power-law activation function acting on noisy voltages, where voltage noise is Gaussian with standard deviation $\sigma$: $g\!\left(v\right) = \langle\lfloor v+\sigma z\rfloor^{n}_{+}\rangle_{z\sim{\cal N}\left(0, 1\right)}$. Note that, when $\sigma>0$, due to the noisiness of the activation function (or the equivalent smoothness in \Cref{eq:noisy_nonlinearity}), all patterns in the network are dense, as firing rates are never exactly zero.
\change{In the limit of $\sigma=0$, this activation function becomes} a noiseless rectified power-law that gives rise to sparse patterns\change{, where a fraction of the patterns is exactly zero. We present results for this limit in Appendix~\ref{subsec:sparse-patterns}}.\looseness=-1

In line with previous approaches \cite{gardner1988space,schonsberg2021efficiency}, we begin by formalizing the computational task for an auto-associative memory as the following: given a set of $P$ memory patterns $\mathbf{r}^\mu$ for $\mu=1\ldots P$, find minimal-norm weights such that each memory pattern is a fixed-point of the network dynamics (\Cref{eq:dynamics_r}):
\begin{equation}\label{eq:optW}\begin{array}{c}
\displaystyle \mathbf{W}^* = \mathop{\textrm{argmin}}_{\mathbf{W}} \left\|\mathbf{W}\right\|_\mathrm{F} \\[10pt]
\displaystyle \text{ s.t. } \mathbf{r}^\mu = \ratefunv\!\left(\mathbf{W}\,\mathbf{r}^\mu\right) \; \forall\mu=1\ldots P
\end{array}
\end{equation} 
\change{where $\left\|\cdot\right\|_\mathrm{F}$ is the Frobenius norm.} \change{This convex problem can be solved using off-the-shelf optimizers \cite{Clarabel_2024} (\Cref{subsec:sparse-opt}).}
We consider the additional constraint that there are no self-couplings, i.e.\ the diagonal elements of the connection weight matrix are zero.\looseness=-1

We deviate from previous approaches \cite{hopfield1982neural, amit1985spin, treves1990graded, schonsberg2021efficiency} in two important ways.
First, we not only require memory patterns to be merely fixed points of the dynamics (which will be guaranteed once \Cref{eq:optW} is solved), but we also study the local stability of these fixed points -- this is critical for a well-functioning auto-associative memory if it is to perform memory recall by converging to a memory pattern when started from a state that is near but not identical to it \cite{hopfield1982neural, khona2022attractor}. 
Specifically, we analyze the Jacobian of the dynamics around each memory pattern
\begin{equation}\label{eq:FixedPointJacobian}
    \mathbf{J}_\mu^*=-\mathbf{I}+\mathbf{g'}\!\left(\mathbf{g}^{-1}\!\left(\mathbf{r}^\mu\right)\right)\circ\mathbf{W}^*
\end{equation}
\change{where $\mathbf{a}\circ \mathbf{B}$ denotes the Hadamard product of $\mathbf{1}\,\mathbf{a}^T$ and $\mathbf{B}$, $g'\left(\cdot\right)$ and $g^{-1}\left(\cdot\right)$ denote the derivative and inverse of $g\left(\cdot\right)$, respectively. }The fixed-point at $\mathbf{r}^\mu$ is stable if all eigenvalues of the Jacobian have negative real parts.

Second, unlike previous approaches that considered binary \cite{hopfield1982neural, amit1985spin} or sparse memory patterns \cite{treves1990graded, schonsberg2021efficiency}, here we specifically focus on dense patterns in which firing rates are never exactly zero: $\mathbf{r}^\mu\in\mathbb{\mathbf{R}}_+^N$. Again, to allow a systematic study of how stability depends on the properties of this distribution, we consider here the specific case of a log-normal distribution of neural activities \cite[found ubiquitously in the brain; ][]{buzsaki2014log}, which we parametrize by its coefficient of variation $\textrm{CV}=\sqrt{\left\langle \delta r^2\right\rangle}/\left\langle r\right\rangle$, while fixing its mean at $\left\langle r\right\rangle=1$ (\Cref{fig:capacity}c), where here and in the following $\left\langle\cdot\right\rangle$ and $\delta$ denote averaging across the distribution of memory patterns and computing the deviation from such an average, respectively. For a hard-rectified power law activation function (i.e.\ $\sigma=0$, \Cref{fig:extdata_fig_1}a), \change{making patterns sparse (\Cref{fig:extdata_fig_1}b), }
scaling the mean of memory patterns while keeping their CV constant would cancel in the Jacobian of the dynamics, \Cref{eq:FixedPointJacobian}, thus leaving stability unaffected. For our soft-rectified activation function, this does not hold exactly, but we find the effects on stability to be negligible (\Cref{fig:sm_optimals}). While such patterns are always technically dense, they can approximate sparse distributions with a sufficiently high CV (we call such patterns `sparse-like').

To summarize, we are interested in how the maximal number of patterns $P$ that can be stored as fixed points, and the fraction of these fixed points that are dynamically stable, depend on parameters $\sigma$, $n$, $\theta$, and CV (as well as $f$, the sparseness of patterns, Appendix~\ref{subsec:sparse-patterns}). For obtaining the maximal $P$, we optimize $\mathbf{W}$ following \Cref{eq:optW} \change{for each combination of parameters}. 

\mysec{Mean-field theory for storage capacity}

A replica analysis of the solution for the optimization problem \Cref{eq:optW} follows the Gardner approach \cite{gardner1988space,schonsberg2021efficiency,cohen2022soft},  without assuming specific connection weights as in the Hopfield model \cite{amit1985spin,treves1990graded}.
In Appendix~\ref{subsec:replica-theory}, we derive a mean-field theory for the capacity to store graded, random memory patterns as fixed points of \Cref{eq:dynamics}, without assumptions on the weights. The analysis applies to either dense patterns with a strictly monotonic activation function (\Cref{fig:capacity}a-c) or to sparse patterns with a rectified monotonic activation function (\Cref{fig:extdata_fig_1}a-b). It generalizes previous work in which a rectified-linear activation function and normalized connection weights were assumed \cite{schonsberg2021efficiency}. As usual, our analysis assumes $P, N\to\infty$ with a fixed memory load $\alpha=P/N$ and uses the replica technique, assuming the replica symmetry ansatz \cite{mezard1987spin}. 

\change{Extending previous work \cite{schonsberg2021efficiency}, our theory predicts the following moments for each element of the connection weight matrix (dotted lines in \Cref{fig:capacity}d-f, and g-i, and \Cref{fig:extdata_fig_1}c-d, for dense and sparse patterns, respectively):
\begin{align}
N \, \left\langle W\right\rangle &=\frac{\theta+\left\langle g^{-1}\!\left(r\right)\right\rangle}{\langle r\rangle} 
\label{eq:meanW}\\
N \, \left\langle W^{2}\right\rangle 	&=\frac{\alpha}{1-\alpha}\frac{\left\langle \delta g^{-1}\!\left(r\right)^2\right\rangle}{\left\langle \delta r^2\right\rangle} 
\label{eq:meanSqrW}
\end{align}
In particular, in line with previous results \cite{gardner1988space}, as $\alpha$ approaches a critical level $\alpha_\mathrm{C}$, the theory predicts the L2-norm of weights to diverge (\Cref{fig:capacity}g-i, dotted lines). This means that for $\alpha<\alpha_\mathrm{C}$ all patterns can be stored, as \Cref{eq:optW} is predicted to have a solution with finite weights, while for $\alpha>\alpha_\mathrm{C}$ no patterns can be stored, as \Cref{eq:optW} is predicted to have no solution.
This theoretical $\alpha_\mathrm{C}$ depends only on pattern sparseness (Appendix~\ref{subsec:sparse-patterns}, \Cref{fig:extdata_fig_1}g) and is independent of the details of the activation function (exponent $n$, smoothness $\sigma$, and threshold $\theta$; \Cref{eq:noisy_nonlinearity}) and other pattern statistics (CV; dashed lines in insets of \Cref{fig:capacity}m-o and in \Cref{fig:extdata_fig_1}e-f), thus extending results from neurons with binary \cite{gardner1988space, tsodyks1988enhanced} or rectified-linear activation functions \cite{treves1990graded, schonsberg2021efficiency} (corresponding to $\sigma=0$, $n=1$) to our broader class of activation functions ($\sigma\geq0$, $n>0$). 
Specifically, in the dense case, we find that the theoretical critical capacity is always $\alpha_\mathrm{C} = 1$, where \Cref{eq:meanSqrW} diverges.}

For a subcritical load, $\alpha<\alpha_\mathrm{C}$, the theory can be compared with numerical experiments (\Cref{subsec:numerics}). The 
theory provides a good match for both the mean ($N \, \left\langle W\right\rangle$ from \Cref{eq:meanW}; solid lines in \Cref{fig:capacity}d-f and \Cref{fig:extdata_fig_1}c) and the L2-norm of a row of the connection weight matrix ($\sqrt{N \, \left\langle W^{2}\right\rangle}$ from \Cref{eq:meanSqrW}; solid lines in \Cref{fig:capacity}g-i and \Cref{fig:extdata_fig_1}d). \change{The apparent discrepancies between theory and numerical simulation (e.g.\ \Cref{fig:capacity}f) are due to finite-size effects, which are more pronounced with high CV patterns and when $\alpha$ approaches zero.}

To test if the resulting dynamics might coincide with those assumed by the `Hebbian' approach, we measured the weight matrix asymmetry and the deviation from normality of the dynamics around the fixed points (\Cref{subsec:non-normality}). Both are non-negligible and increase with the memory load $\alpha$ and pattern variation (\Cref{fig:capacity}j-l), indicating that the resulting dynamics are distinct from those of energy-based models, and are similar to those previously found in attractor networks with optimized weights \cite{stroud2023optimal}.\looseness=-1

\change{The theory also correctly predicts the phase transition in the fraction of successfully stored patterns, and that the location of this phase transition, the critical load $\alpha_\mathrm{C}$, is independent of parameters beyond sparseness (solid lines in insets of \Cref{fig:capacity}m-o, and in \Cref{fig:extdata_fig_1}e, and black and purple lines in \Cref{fig:extdata_fig_1}f).}

Importantly, investigating the emergent stability of patterns in the subcritical regime reveals a second phase transition: all patterns tend to be stable up to a new critical value $\alpha_\mathrm{S}<\alpha_\mathrm{C}$, above which all patterns tend to be unstable (\change{solid lines in main plots of} \Cref{fig:capacity}m-o, and in \Cref{fig:extdata_fig_2}a-b). \change{While the first phase transition is the well-known capacity phase transition described for both `Hebbian' \cite{amit1985spin,kanter1987associative,tsodyks1988enhanced,treves1990graded} and `Gardner' \cite{gardner1988space, schonsberg2021efficiency} settings, the stability phase transition is unique to our analysis (but is known in some random dynamical systems, see Discussion).} In the following, we investigate this phenomenon more closely.\looseness=-1

\mysec{Theory of stability for dense patterns}
For dense patterns, $\alpha_\mathrm{C}=1$ and for any $\alpha<\alpha_\mathrm{C}$ the solution to \Cref{eq:optW} is given in closed-form \change{in terms of the pseudo-inverse $\mathbf{R^\dagger}$}(\Cref{subsec:kkt}):
\begin{equation}\label{eq:denseW}
\mathbf{W}^*=\mathbf{V} \, \mathbf{R^\dagger}=\mathbf{V} \, 
\left(\mathbf{R}^\mathsf{T}\, \mathbf{R}\right)^{-1} \, \mathbf{R}^\mathsf{T}
\end{equation}
where $\mathbf{R},\mathbf{V}\in\mathbb{\mathbb{R}}^{N\times P}$ are the stored memory patterns $R_{i\mu}=r_i^{\mu}$ and $V_{i\mu}=g^{-1}(R_{i\mu})+\theta$. A slightly more complicated expression is available when avoiding self-couplings, \Cref{eq:optWzeroDiagonal}, but we find that we can neglect this difference and use \Cref{eq:denseW} in the derivation of stability. This generalizes earlier results derived for storing binary patterns, in which case weights defined through the patterns' pseudo-inverse were already proposed in the `Hebbian' approach \cite{personnaz1985information, kanter1987associative}. However, those results were limited to a linear activation function $\mathbf{V}=\mathbf{R}$, rendering $\mathbf{W}^*$ symmetric (and thus allowing the construction of an energy function). In the general case of a non-linear activation function, $\mathbf{W}^*$ obtained from \Cref{eq:denseW} is not symmetric and thus falls outside the scope of the `Hebbian' approach.\looseness=-1

\begin{figure*}
\centering
\includegraphics[width=\textwidth]{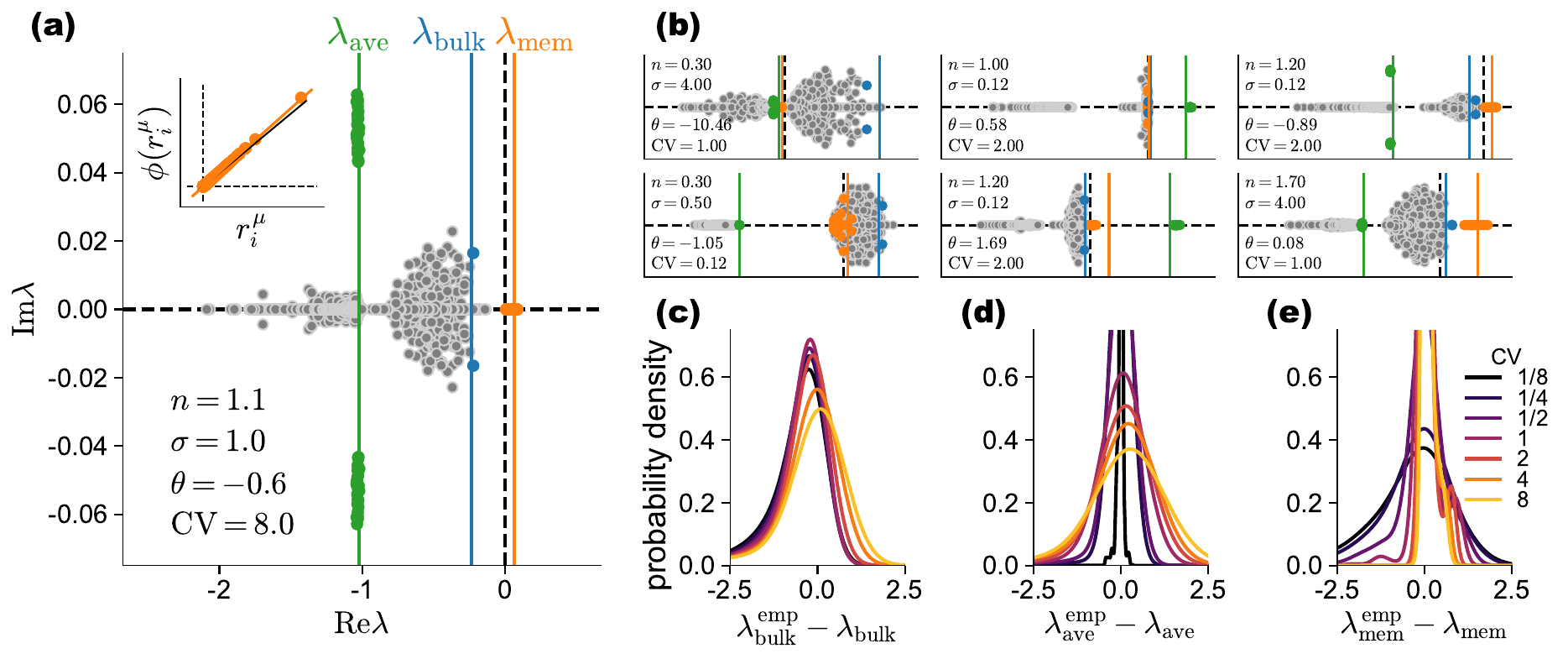}
\caption{\label{fig:spectrum}
\textbf{Fixed-point stability.} (a) Example spectrum -- eigenvalues of \change{$P=32$} Jacobians $\mathbf{J_\mu}$ of a single network (\change{$N=256$, } other parameters are shown in the legend). Vertical lines show the theoretical values of $\lambda_\mathrm{bulk}$ (blue), $\lambda_{\mathrm{ave}}$ (green), and $\lambda_{\mathrm{mem}}$ (orange). The corresponding empirical eigenvalues are marked with matching colors: the largest eigenvalue of the bulk (blue dot), the eigenvalue associated with the Jacobian average (green dot), and the eigenvalue associated with the memory pattern (orange dot). Inset shows a near-linear relationship between $r^\mu_i$ (x-axis) and $\phi(r^\mu_i)$ (y-axis), a prerequisite for an outlier $\lambda_\mathrm{mem}$. (b) Same as (a), for six different parameter sets (shown in the legends). \change{The three columns respectively show illustrative examples for which $\lambda_\mathrm{bulk}$ (left), $\lambda_{\mathrm{ave}}$ (middle), or $\lambda_{\mathrm{mem}}$ has the largest real part.} (c-e) Probability densities of the difference between the empirical and predicted value of $\lambda_\mathrm{bulk}$ (c), $\lambda_\mathrm{ave}$ (d), and  $\lambda_\mathrm{mem}$ (e) \change{at different CV values (color coded) while varying other parameters. For each CV value, we simulated a network of size $N=128$ for each of the $6,400$ combinations of the following parameter values: $n\in\left\{0.1\,i:i=1\dots20\right\}$, $\theta\in\left\{-4, -2, -1, 0, 1\right\}$, $\sigma\in\left\{0,1/8,1/4,1/2,1,2,4,8\right\}$, and $P\in\left\{\left\lfloor0.1\,N\,i\right\rfloor:i=1\dots8\right\}$. For each parameter combination, optimal weights $\mathbf{W}$ were obtained using \Cref{eq:denseW}. Parameter combinations were omitted from subsequent analyses if they did not result in satisfying the following constraints: $\left|\lambda_\mathrm{bulk}\right|<10$ (c), $\left|\lambda_\mathrm{ave}\right|<10$ (d), $\tau_\mathrm{mem}>0.95$ (e).}}


\end{figure*}

To characterize $\alpha_\mathrm{S}$, the critical load for the stability of $\mathbf{J}_\mu^*$, we note that from \Cref{eq:FixedPointJacobian,eq:denseW}:
\begin{equation}\label{eq:JacobianOpt}
    \mathbf{J}_\mu^*=-\mathbf{I}+\underbrace{\mathbf{g'}\!\left(\mathbf{g}^{-1}\!\left(\mathbf{r}^\mu\right)\right)\circ\mathbf{V}}_{\displaystyle \bar{\mathbf{V}}_\mu} \, \mathbf{R}^\dagger
\end{equation}
and build on our previous analysis of the eigenvalue spectrum of random matrices $
\mathbf{M}=-\mathbf{I}+\mathbf{X}\,\mathbf{Y}^\dagger$ when pairs of the corresponding entries of $\mathbf{X}$ and $\mathbf{Y}$ are (jointly) independent and identically distributed (i.i.d.), correlated Gaussian, and $\alpha$ simply denotes the dimension-ratio (width/height) of $\mathbf{X}$ and $\mathbf{Y}$ (a generalization of $\alpha$ being the load in our case) \cite{cohen2025eigenvalue}. 
This analysis provides an exact result for the support of the eigenvalue spectrum (\Cref{fig:SM_GaussianPseudoinverse}) and the largest real eigenvalue $\lambda_\mathrm{bulk}$ of $\mathbf{M}$ (\Cref{subsec:lambda-bulk}, \Cref{eq:GaussianPseudoinverseLambdaSupport}). 
Interestingly, we find that, just as the Jacobians of the original network, $\mathbf{J}_\mu^*$, the Jacobians defined by $\mathbf{M}$ (for any choice of $\mathbf{X}$ and $\mathbf{Y}$) also undergo a phase transition: they are either all stable (for $\alpha<\alpha_\mathrm{S}$) or all unstable (for $\alpha>\alpha_\mathrm{S}$)
(\Cref{subsec:lambda-bulk}, \Cref{eq:GaussianPseudoinverseStability}). 
To apply the general theory of the eigenvalue spectrum of $\mathbf{M}$ to the eigenvalue spectrum of $\mathbf{J}_\mu^*$ (\Cref{eq:JacobianOpt}), we choose $\mathbf{X}$ and $\mathbf{Y}$ such that the variances and covariance of 
$X_{i\mu}$ and $Y_{i\mu}$ are respectively  $c_\mathrm{ff} = \langle \delta f\!\left(r,r'\right)^2\rangle$, $c_\mathrm{rr} = \langle \delta r^2\rangle$, and $c_\mathrm{rf} = \left\langle \delta r\, \delta f\!\left(r,r'\right)\right\rangle$, where $f\!\left(r,r'\right)= g'\!\left(g^{-1}\!\left(r'\right)\right)\,\left(g^{-1}\!\left(r\right)+\theta\right)$ for independent variables $r$, $r'$. With these substitutions,
\begin{align}\label{eq:lambda_bulk}
    \lambda_\mathrm{bulk} 
    &= -1 + \frac{c_\mathrm{rf}}{c_\mathrm{rr}}+ \frac{1}{c_\mathrm{rr}}\sqrt{\frac{\alpha}{1-\alpha} \, \left(c_\mathrm{rr}\,c_\mathrm{ff}-c^2_\mathrm{rf}\right)} 
\end{align}
Solving for the largest value of $\alpha$ for which $\lambda_\mathrm{bulk}<0$, the resulting critical load for stability from this analysis is:
\begin{equation}\label{eq:alphaSbulk}
\alpha_{\mathrm{S}}^{\mathrm{bulk}} = \frac{\max\left(0,c_\mathrm{rr}-c_\mathrm{rf}\right)^2}{c_\mathrm{rr}\,c_\mathrm{ff}-c_\mathrm{rf}^2+\left(c_\mathrm{rr}-c_\mathrm{rf}\right)^2}
\end{equation}

The foregoing stability analysis was based on the zero-crossing of the `rightmost point' of the bulk of the eigenvalue spectrum of $\mathbf{M}$, $\lambda_\mathrm{bulk}$. However, we note that this eigenvalue spectrum can only be an approximation to that of $\mathbf{J}_\mu^*$ (\Cref{fig:spectrum}a), since the entries of $\mathbf{J}_\mu^*$ are neither i.i.d.\ (due to the $g'\!\left(g^{-1}\!\left(\mathbf{r^\mu}\right)\right)$ term in $\bar{\mathbf{V}}_\mu$, which couples different rows), nor Gaussian (since $\mathbf{V}$ is a deterministic function of $\mathbf{R}$, rather than being a jointly distributed random variable). Despite these facts, we empirically find that the eigenvalue spectrum of $\mathbf{M}$ often provides an acceptable approximation to at least the bulk of the eigenvalue spectrum of $\mathbf{J}_\mu^*$, so that $\lambda_\mathrm{bulk}$ often predicts stability well (\Cref{fig:spectrum}a,c and \Cref{fig:spectrum}b, left column). Nevertheless, there are also cases when $\lambda_\mathrm{bulk}$ alone is an imperfect predictor of stability (\Cref{fig:spectrum}b, middle and right columns). Thus, we analyze two specific outlier eigenvalues that \change{we assume might be responsible for instability in most cases when $\lambda_\mathrm{bulk}$ alone would predict stability}.\looseness=-1

The analysis of each of the two outlier eigenvalues is based on assuming the existence of a specific eigenvector of $\mathbf{J}_\mu^*$ and computing its corresponding eigenvalue, which may be an outlier.
In the first case, we make the assumption that the uniform vector, $\boldsymbol{1}$, is an eigenvector. If such an eigenvector existed, the corresponding eigenvalue, $\lambda_\mathrm{ave}^\mu$, would be the (average) row-sum of $\mathbf{J}_\mu^*$, $\bar{J}$. Thus, using replica theory (\Cref{eq:meanW}) to compute $N \, \langle W \rangle$, the expectation of this outlier eigenvalue is:
\begin{equation}
\label{eq:lambdaAve}
\lambda_\mathrm{ave}=\left\langle \lambda_\mathrm{ave}^\mu \right\rangle = \left\langle \bar{J}\right\rangle 
    =-1+\left\langle g'\!\left(g^{-1}\!\left( r\right)\right) \right\rangle \, \frac{\theta+\left\langle g^{-1}\!\left(r\right)\right\rangle}{\langle r\rangle}
\end{equation}
where by taking the expectation over $g'\!\left(g^{-1}\!\left(\mathbf{r}^{\mu}\right)\right)$, we have again ignored its memory pattern-dependence.
\change{Despite this approximation, and} even though the uniform vector is not an eigenvector of $\mathbf{J}_\mu^*$ in general, when we empirically identify the closest eigenvector (\Cref{subsec:lambda-ave}), we find that the associated eigenvalue is often close to $\lambda_\mathrm{ave}$ (\Cref{fig:spectrum}a,d), and can determine the dynamical stability of $\mathbf{J}_\mu^*$ (\Cref{fig:spectrum}b, middle column). 

In the second case, we assume that there exists some constant $c$ for which $\mathbf{r}^{\mu}+c\,\boldsymbol{1}$ is an eigenvector of $\mathbf{J}_\mu^*$. In this case, the following equation must hold (\Cref{subsec:lambda-mem}):
\begin{equation} \label{eq:memoryEigenvector}\boldsymbol{\phi}\!\left(\mathbf{r}^{\mu}\right)=\left(\lambda_{\mathrm{mem}}^{\mu}+1\right) \, \mathbf{r}^{\mu}+c \, \left(\lambda_{\mathrm{mem}}^{\mu}-\lambda_{\mathrm{ave}}\right) \, \boldsymbol{1}
\end{equation}
where $\phi_i\!\left(\mathbf{\mathbf{r}}\right) = \phi\!\left(r_i\right) = f\!\left(r_i,r_i\right)$.
In practice, \Cref{eq:memoryEigenvector} may not hold exactly, so more generally we can characterize the (approximately) linear relationship between $\mathbf{r}^\mu$ and $\boldsymbol{\phi}\!\left(\mathbf{r}^\mu\right)$ by the expectation of its correlation coefficient, $\tau_{\mathrm{mem}}^{\mu}$, and slope, $\lambda_{\mathrm{mem}}^{\mu}$:
\begin{align}
    \tau_{\mathrm{mem}} &= \left\langle\tau_{\mathrm{mem}}^{\mu} \right\rangle = c_\mathrm{r\phi}/\sqrt{c_\mathrm{rr}\,c_\mathrm{\phi\phi}}\label{eq:tauMem}\\
    \lambda_{\mathrm{mem}} &= \left\langle \lambda_{\mathrm{mem}}^{\mu} \right\rangle = -1+c_\mathrm{r\phi}/c_\mathrm{rr}\label{eq:lambdaMem}
\end{align}
where $c_\mathrm{r\phi} =\left\langle \delta r \, \delta \phi\!\left(r\right)\right\rangle$, $c_\mathrm{\phi\phi} =\left\langle \delta \phi\!\left(r\right)^2\right\rangle$ are additional (co)variance terms, and the assumption on which the analysis of $\lambda_{\mathrm{mem}}$ is based can be expressed as $\tau_{\mathrm{mem}}\approx1$. 
To empirically test the relevance of  $\lambda_{\mathrm{mem}}^{\mu}$ (\Cref{eq:lambdaMem}) for determining the stability of $\mathbf{J}_\mu^*$, it is straightforward to identify the eigenvector of $\mathbf{J}_\mu^*$ associated with the highest $\tau_{\mathrm{mem}}^{\mu}$, and compare the corresponding eigenvalue to $\lambda_\mathrm{mem}$ (\Cref{fig:spectrum}a,e). Indeed, we find cases where  $\lambda_\mathrm{mem}^{\mu}$ determines the dynamical stability of $\mathbf{J}_\mu^*$ (\Cref{fig:spectrum}b, right column).

For characterizing the transition from the stable to the unstable regime, we analyze the case when exactly half of the patterns stored in a network are stable\change{, which is captured by }the median $\lambda_\mathrm{ave}^\mu$ and $\lambda_\mathrm{mem}^\mu$ across patterns. \change{Assuming that the zero-crossing of this median is well approximated by the zero-crossing of the mean value, stability is determined by the zero crossings of the expectations we computed above} (\Cref{eq:lambdaAve,eq:lambdaMem}). As neither of these \change{theoretical} expectations
depend on the number of stored patterns, we can estimate the critical load for stability \change{using the Heaviside step function $\Theta$ }as:
\begin{equation}\label{eq:alphaS}\alpha_{\mathrm{S}}=\alpha_{\mathrm{S}}^{\mathrm{bulk}} \, \Theta\!\left(-\lambda_\mathrm{ave}\right) \,\left[1-\Theta\!\left(\lambda_\mathrm{mem}\right) \,  \Theta\!\left(\tau_\mathrm{mem}-0.95\right)\right]
\end{equation}
where $\tau_\mathrm{mem}>0.95$ is an arbitrary choice for the domain where we consider the prediction about $\lambda_\mathrm{mem}$ to apply. 

We first focus on the zero crossing of $\alpha_{\mathrm{S}}$ as a function of the parameters, i.e.\ the parameter regime where stability is possible at all. For this, we analyze the zero crossing of each of the three terms of \Cref{eq:alphaS} separately by solving the corresponding inequalities: $\alpha_\mathrm{S}^\mathrm{bulk}>0$, $\lambda_\mathrm{ave}<0$, and $\lambda_\mathrm{mem}<0$ (assuming $\tau_\mathrm{mem}>0.95$). The dependence of each of these zero crossings on $\theta$ is relatively straightforward. The zero crossing of $\alpha_\mathrm{S}^\mathrm{bulk}$ does not depend on $\theta$ (although the exact value of $\alpha_\mathrm{S}^\mathrm{bulk}$ when it is greater than zero does, \Cref{subsec:threshold}). Both $\lambda_\mathrm{ave}$ and $\lambda_\mathrm{mem}$ depend on $\theta$ linearly (\Cref{eq:lambdaAve}, and \Cref{subsec:threshold}), and so the condition for their zero crossing can be expressed respectively as 
\begin{align}
\theta_0^\mathrm{ave}&=\frac{\langle r\rangle}{\left\langle d\!\left(r\right) \right\rangle} - \left\langle g^{-1}\!\left(r\right)\right\rangle \label{eq:theta_lambdaAve}\\
\text{and }
\theta_0^\mathrm{mem}&=\frac{\left\langle \delta r\, \delta r\right\rangle - \left\langle \delta r \, \delta d\!\left(r\right)\, g^{-1}\!\left(r\right)\right\rangle}{\left\langle \delta r \, \delta d\!\left(r\right)\right\rangle} \label{eq:theta_lambdaMem}
\end{align}
where $d\!\left(r\right)=g'\!\left(g^{-1}\!\left(r\right)\right)$, and the RHSs of \Cref{eq:theta_lambdaAve,eq:theta_lambdaMem} implicitly depend on the other parameters. 

\begin{figure*}
\centering
\includegraphics[width=\linewidth]{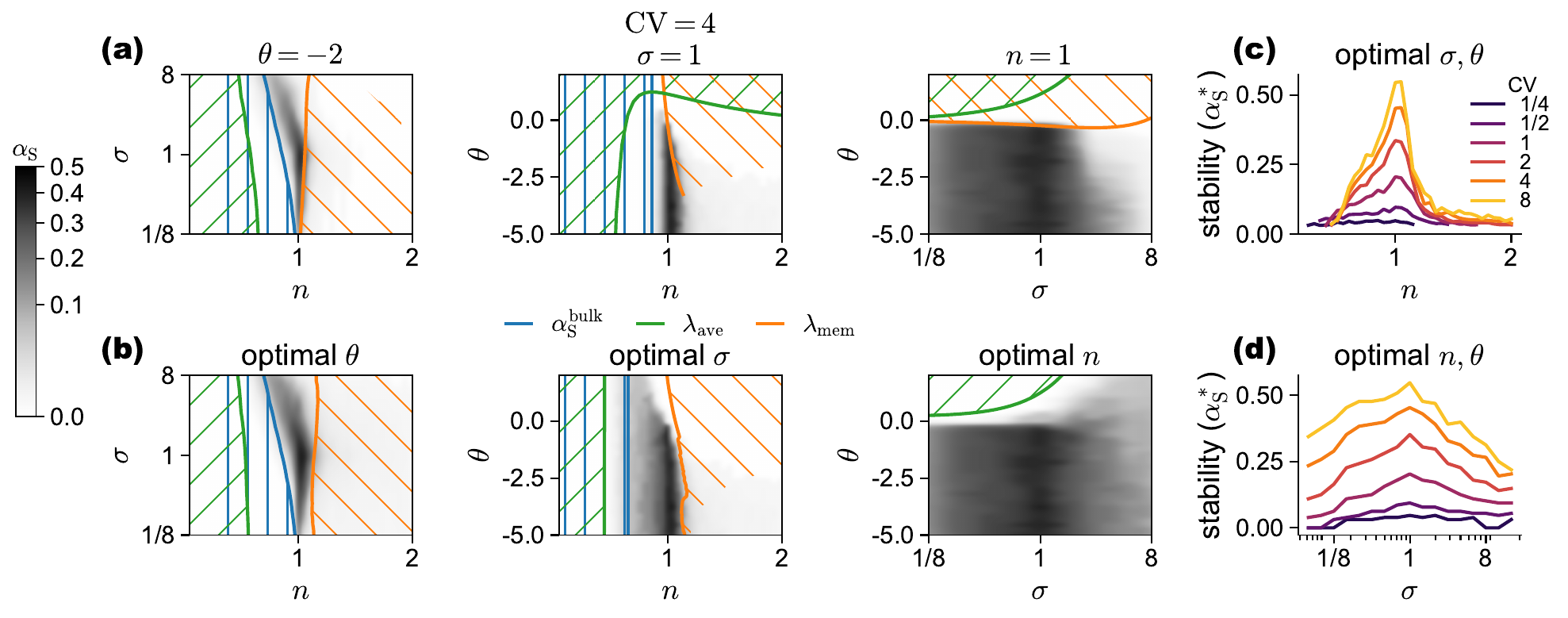}
\caption{\label{fig:phase-diagram}
\textbf{Phase diagrams for the critical load for stability.} (a-b) Numerical simulation results for the critical load for stability, $\alpha_\mathrm{S}$ (grayscale map), using different choices of activation function exponent, $n$, smoothness, $\sigma$, and threshold, $\theta$, along with theoretical predictions for the different terms controlling $\alpha_\mathrm{S}$ according to \Cref{eq:alphaS}: \change{hatched areas indicate regions where $\alpha^\mathrm{bulk}_\mathrm{S}=0$ (blue), $\lambda_\mathrm{ave}>0$ (green), and $\lambda_\mathrm{mem}>0$ (orange); thick colored lines indicate boundaries of the corresponding regions.} $\lambda_\mathrm{mem}$ is considered only in the domain where $\tau_\mathrm{mem}>0.95$ \change{(i.e., regions with missing orange hatching have $\tau_\mathrm{mem}<0.95$)}.
In (a), two parameters are varied (axes), while the third parameter (title) is kept fixed. In (b), the third parameter is optimized for each combination of the two parameters that are being varied. $\mathrm{CV}=4$ in all cases shown.
(c) The maximal value of $\alpha_\mathrm{S}$ for each value of $n$ (x-axis) and $\mathrm{CV}$ (color coded). (\change{Numerical simulations,} parameters $\sigma$ and $\theta$ were optimized for each combination of $n$ and $\mathrm{CV}$.) 
(d) The maximal value of $\alpha_\mathrm{S}$ for each value of $\sigma$ (x-axis) and $\mathrm{CV}$ (color coded as in (c)).  (\change{Numerical simulations,} parameters $n$ and $\theta$ are optimized for each combination of $n$ and $\mathrm{CV}$.) \change{Numerical values of $\alpha_\mathrm{S}$ reported in (a-d) were computed based on simulations using a fixed set of parameter combinations. These combinations were generated using a fixed grid of $\mathrm{CV}$ values (see legend in (c)), and exponents $n\in\left\{0.05\,i:i=1\dots40\right\}$, and an adaptive grid of $\sigma$ and $\theta$ values. For each parameter combination, optimal weights $\mathbf{W}$ were obtained using \Cref{eq:denseW} for a network of size $N=128$. We started from $P=4$, $\sigma\in\left[1/16, 16\right]$ (with equal steps in log scale), and $\theta\in\left[-6,3\right]$ (with equal steps in linear scale), and at each iteration increased $P$ by $1$ and made the range of $\sigma$ and $\theta$ narrower, removing leading or tailing values with zero stable patterns and re-adjusting grid steps. The numerical value of $\alpha_\mathrm{S}$ reported for a given parameter combination was defined as $P/N$ for the maximal value of $P$ for which at least half of the patterns were dynamically stable.}
}
\end{figure*}

\mysec{Phase diagrams}
In order to gain insights into the complex dependence of the zero crossings of the three terms of \Cref{eq:alphaS} on the combination of parameters (pattern statistics $\textrm{CV}$, activation function smoothness $\sigma$, exponent $n$, and threshold $\theta$), we solved the corresponding equations systematically over a grid of parameter combinations, and compared these predictions with numerically obtained values of $\alpha_\mathrm{S}$.
\Cref{fig:phase-diagram}a presents the resulting phase diagrams of $\alpha_\mathrm{S}$ for different pairs of parameters when the third parameter is kept fixed, showing where $\alpha_\mathrm{S}^\mathrm{bulk}$, $\lambda_\mathrm{ave}$, and $\lambda_\mathrm{mem}$ (the three main determinants of $\alpha_{\mathrm{S}}$ according to \Cref{eq:alphaS}) are predicted to cross zero (blue, green, and orange lines, respectively) as well as the numerically computed values of $\alpha_{\mathrm{S}}$ (grayscale map). We note that $\lambda_\mathrm{ave}$ constrains stability only for $\theta>1, n\approx1$ (\Cref{fig:phase-diagram}a middle panel), while the upper bound on the threshold $\theta$ is due to the zero crossing of $\lambda_\mathrm{mem}$ (\Cref{eq:theta_lambdaMem}). 
The superior stability near  $n=1$ is explained by noting that while $\lambda_\mathrm{bulk}$ prevents stability in most of the sublinear regime, where $c_\mathrm{rf}\ge c_\mathrm{rr}$, $\lambda_{\mathrm{mem}}$ prevents stability in most of the supralinear regime, where $c_\mathrm{r\phi}\ge c_\mathrm{rr}$ (interestingly, the former is derived by neglecting the pattern-specific statistical dependency between the rows of the Jacobian, while the latter is a result of this dependency)\looseness=-1.

\change{To gain a better intuition into the relationship between the exponent of the activation function, $n$, and putative sources of fixed-point instability, we conducted numerical experiments. We embedded activity patterns as fixed-points of a network (using \Cref{eq:optW}), initialized neural activities with a perturbed version of one of those patterns, and measured if the initial dynamics (at time $0$) drove the activities towards or away from the corresponding fixed point. We observed that perturbing the overall scale of the pattern (without affecting its direction) caused instability in the supralinear, but not in the linear or sublinear regimes (\Cref{fig:sm_intuition}, top). Conversely, perturbing the direction of activities (without affecting their overall scale) caused comparatively more instability in the sublinear regime than in the linear or supralinear regimes (\Cref{fig:sm_intuition}, bottom). Scaling activity as a source of instability in the supralinear regime is related to $\lambda_\mathrm{mem}$ being the cause of instability in our theoretical analysis (\Cref{fig:phase-diagram}a-b, orange).
The intuition for this can be made explicit for $\theta=0$ and $\sigma=0$, as in this case the Jacobian (\Cref{eq:FixedPointJacobian}) at a fixed point $\mathbf{r}=[\mathbf{v}]^n$ is $\mathbf{J}=-\mathbf{I}+ n \, [\mathbf{v}]^{n-1} \circ \mathbf{W}$ and, as $\mathbf{W} \, \mathbf{r}=\mathbf{v}$, we thus have $\mathbf{J} \, \mathbf{r} =-\mathbf{r}+n \, [\mathbf{v}]^{n-1} \circ \mathbf{v} = \left(n-1\right) \, \mathbf{r}$. Therefore, in this case, the fixed-point $\mathbf{r}$ is also an eigenvector of the Jacobian and is unstable for $n>1$. Conversely, in the sublinear regime, the effect of direction perturbations is amplified for small values due to the large slope of nonlinearity near zero. Thus, these kinds of perturbations may be related to $\lambda_\mathrm{bulk}$ as a source of instability in the theory, which is consistent with $\alpha_\mathrm{S}^\mathrm{bulk}=0$ in most of the sublinear regime (\Cref{fig:phase-diagram}a-b, blue).}

We can now also better understand the limitations of the theory: the zero crossing of $\lambda_{\mathrm{mem}}$ yields a discontinuous fall of $\alpha_\mathrm{S}$ from a finite value to $0$ (\Cref{eq:alphaS}), and in many cases the maximum is achieved at this boundary. Thus, deviations of the empirical value of this outlier would cause large prediction errors at that boundary. Furthermore, in two regimes -- large $\sigma$ and low $\theta$ -- we observe regions where the assumption on the memory-pattern related outlier is violated \change{(see missing hatchings in \Cref{fig:phase-diagram}a)}, and the empirical value of the outlier is larger than theory predicts, so the empirical $\alpha_\mathrm{S}$ might be zero while theory predicts otherwise.

To complement these results, \Cref{fig:phase-diagram}b presents similar phase diagrams, but where the third parameter is optimized for each pair of systematically varied parameters, rather than kept fixed. The results remain qualitatively similar to those obtained above (\Cref{fig:phase-diagram}a). This view highlights the superiority of $n=1$ and $\sigma=1$ in terms of the resulting number of stable memories, which is further demonstrated across all $\textrm{CV}$ values in \Cref{fig:phase-diagram}c-d for $n$ and $\sigma$, respectively. On the other hand, the effect of $\theta$ seems minor, as long as it is below an upper bound. 

\mysec{Optimal parameters}
\Cref{fig:sm_optimals} presents the maximal achievable value for the stability critical load as a function of pattern statistics, as well as the optimal parameters where the optimum is achieved for various choices of $\left<r\right>$. 
Our theory predicts improvement of stability with pattern variation (\Cref{fig:phase-diagram}c-d, \Cref{fig:sm_optimals}a). Those high-variance memory patterns are sparse-like, with many near-zero values and few large activations, so the improved capacity we report for those echoes the benefit of sparseness in our results (\Cref{fig:extdata_fig_2}a,e) and previous works \cite{treves1990graded, tsodyks1988enhanced}. Such sparse-like patterns are also consistent with the known phenomenology of neurons in the brain 
\cite{willmore2001characterizing, olshausen2004sparse}.
As noted above, such patterns require highly asymmetric weights and lead to dynamics distinct from those of energy-based models (\Cref{fig:capacity}l).

Our analysis predicts substantial stability only in the near-linear region $n\approx1$ (\Cref{fig:phase-diagram}c, \Cref{fig:sm_optimals}b). The benefits of a threshold-linear activation function are also observable for sparse patterns (\Cref{fig:extdata_fig_2}e), where it is optimal for $\textrm{CV}\ge1$. This is consistent with previous literature on the storage of graded memory patterns using the `Hebbian' approach, which focused on this activation function \cite{treves1990graded, schonsberg2021efficiency} and with the machine learning literature using similar activations \cite{dubey2022activation}, but contrasts with other approaches that advocate for the benefits of a supralinear activation function \cite{rubin2015stabilized, ahmadian2021dynamical}. 

Numerical results demonstrate that the optimal choice of $\sigma$ is a finite value (\Cref{fig:phase-diagram}d, \Cref{fig:sm_optimals}c). This value is not well predicted by theory due to the abovementioned imprecision in the estimation of the zero crossing of $\lambda_\mathrm{mem}$.

\change{Our analysis predicts a combination of a negative threshold ($\theta<0$) and net negative recurrent weights ($m<0$) to be optimal, as shown for the near-optimal values $n=1,\sigma=1$ in \Cref{fig:capacity}f. This implies that neurons are biased toward activity ($\theta<0$) and are `held back' by recurrent connections which are dominated by inhibition ($m<0$). The optimality of the negative threshold is supported by our theory of stability, bounding the threshold from above (\Cref{fig:phase-diagram}a-b), and is evident empirically at the optimal parameters for dense patterns (\Cref{fig:sm_optimals}d), as well as for sparse patterns (\Cref{fig:extdata_fig_2}c). \change{While this result may seem counterintuitive, as a negative threshold with fixed weights would diminish pattern sparseness, note that our analysis considers networks with optimized (and thus threshold-dependent) weights.} Such a negative threshold was reported as a condition for stability in some other, unrelated models \cite{golomb1990willshaw, ben1995theory, barak2021mapping}\change{, as well as a similar rate model with random weights \cite{kadmon2015transition}}. 
Due to the relationship between average weight and threshold in our theory (\Cref{eq:meanW}), a negative threshold implies that the average weight also tends to be negative.
In contrast, previous theories of attractor models using standard capacity considerations, without analyzing stability, were only able to predict a balance between the threshold and the average weight (\Cref{eq:meanW}), without constraining the sign of either individually \cite{schonsberg2021efficiency}.}\looseness=-1

\mysec{Discussion\label{sec:discussion}}

Our analysis highlights the conditions for memory pattern stability in attractor networks performing auto-associative memory recall. The resulting theory \change{identifies} a \textit{stability phase transition} at a number of patterns which is proportional to the number of neurons, below the \textit{capacity phase transition}, which does not account for stability. 
\change{Albeit a direct experimental test of these phase transitions may not be possible, our theory makes predictions about the properties of neural circuits that maximize the number of patterns at which they occur. Specifically, it} constrains the range of possible choices for single-neuron activation functions and establishes that near-linear activation, a negative threshold, and non-zero noise levels are optimal for the stable recall of dense memory patterns.

The predictions above are relevant to neuroscience and cannot be derived from the non-biological, energy-based auto-associative memory networks commonly used in the field. Our work thus paves the way for designing networks with biologically relevant dynamics capable of storing a large number of patterns as stable fixed points by standard optimization approaches. In particular, optimization only needs to consider the existence of fixed points, not their stability, as the latter is guaranteed under the stability phase transition. 
Future research will need to establish if biologically plausible local learning rules, rather than optimization, can achieve similar results.

\change{Our analyses focused on discrete attractor networks in which stable fixed points (0-dimensional attractors) correspond to distinct (episodic) memories \cite{hopfield1982neural,tsodyks1988enhanced,treves1990graded,krotov2016dense}. In contrast, a large body of influential work models various cognitive processes, including eye movement control \cite{seung1996brain}, parametric working memory \cite{compte2000synaptic}, spatial navigation \cite{skaggs1994model, burak2009accurate}, and evidence accumulation \cite{wong2006recurrent}, using recurrent neural networks that instead store manifold (e.g., ring) attractors (i.e., networks for which there exist 1- or higher dimensional manifolds within which dynamics are stable or quasi-stable). However, given the additional complexity of manifold attractors, the analysis of capacity and stability in these networks has mostly remained limited to the stability of a single manifold \cite{pollock2020engineering,darshan2022learning,sagodi2024back,schonsberg2024continuous,clark2025symmetries}, or to the capacity for multiple attractors without considering their stability \cite{battaglia1998attractor}. 
When the stability of multiple attractors has been studied, it was for networks with binary neurons \cite{monasson2013crosstalk, battista2020capacity}. 
While many of these works analyzed the within-manifold stability of dynamics \cite[see also][for multiple attractors]{agmon2023simultaneous}, in the above we only consider analyses of the out-of-manifold stability of the dynamics as analogous to our stability of the dynamics around fixed points in terms of limiting the number of attractors that can be simultaneously stored in the network. 
Interestingly, despite only analyzing the stability of a single ring attractor, one earlier work also found that stability to perturbations proportional to stored patterns (the equivalent of our $\lambda_\mathrm{mem}$) needs to be analyzed in addition to stability due to `generic' perturbations (the equivalent of our $\lambda_\mathrm{bulk}$) \cite{darshan2022learning}. However, in this case the latter kind of stability was analyzed for random perturbations to weights rather than due to weights being optimized for a large number of additional attractors, and thus the resulting mathematical ensemble was more trivial.
Therefore, none of this previous work made predictions about the properties of circuits that would allow them to maximize the number of stable (manifold) attractors they are able to store. 
}

Several areas of science study dynamical systems where fixed points and their dynamic stability may be of interest \cite{may1972will}. When theoretical analyses were conducted, similar to our approach, they focused on the eigenvalue spectra of the dynamics' Jacobians at fixed points. A full characterization of the spectrum is possible for a variety of random matrix ensembles \cite{sommers1988spectrum,guhr1998random,mehta2004random,bai2010spectral,stern2014dynamics,ahmadian2015properties,poley2024eigenvalue} and exhibits a universality property \cite{tao2009random}.
Many random dynamical systems (whose local linearization induces a random matrix ensemble) exhibit a stability phase transition as some parameter of the ensemble is increased
\cite{may1972will,sompolinsky1988chaos,ahmadian2015properties,kadmon2015transition,fyodorov2016nonlinear}. Such random dynamical systems can have an exponential number of fixed points, with stability that depends on a global criterion \cite{wainrib2013topological,stern2014dynamics,ben2021counting}.
Here, we were able to go beyond random matrices \cite[or low-rank perturbations thereof;][]{orourke2014low, mastrogiuseppe2018linking} to describe fixed-point stability in a system resulting from an optimization process, where the connection weight matrix (and subsequently the Jacobian) has considerable structure defined by the objective function it minimizes. The stability of fixed points in an optimized network was previously described analytically only for simple cases, when storing a single pattern of activity by optimizing the weights of a single feedback neuron \cite[i.e.\ a rank-1 perturbation of the weights; ][]{rivkind2017local}, or a very small number of patterns using a rank-2 perturbation \cite{schuessler2020dynamics}. Our work provides conditions for the dynamical stability of networks optimized to store a large number of memory patterns.\looseness=-1 

Our analysis of the eigenvalue spectrum of the Jacobian may be applicable to many other systems where an optimal solution is given by a pseudo-inverse, as in multivariate linear regression problems $S^*=\arg\min_{S} \left\|A-B\,S\right\|_\mathrm{F}$ for rectangular matrices $A$, $B$ and a square matrix $S$. Those arise naturally in many areas of science, such as in the case of time series analysis where $A$ and $B$ are time-lagged ``Trajectory Matrices'', useful for the analysis of non-linear dynamics \cite[in ``Singular Spectrum Analysis''; ][]{broomhead1986extracting, golyandina2001analysis}, for Multivariate Calibration in Chemometrics \cite[in ``Principal Component Regression''; ][]{martens1992multivariate} and in the study of fluid dynamics  \cite[in ``Dynamic Mode Decomposition''; ][]{schmid2022dynamic}. Our approach, in particular deriving expressions similar to \Cref{eq:GaussianPseudoinverseStability} with appropriate assumptions, should allow the analysis of the optimal solution's spectrum and its stability in this broad range of problems.\looseness=-1

\mysec{Materials and methods}
\mysubsec{Optimal weights for sparse patterns\label{subsec:sparse-opt}}
\change{In the sparse case, \Cref{eq:optW} can be numerically solved as quadratic optimization under $PN$ equality and inequality constraints: 
\begin{align*}
\mathbf{W}^{*} = \arg\min\mathbf{W}\ s.t.\ \forall\mu\in1\dots P,\forall k\in1\dots N \\
\begin{cases}
    \sum_j W_{kj} r^\mu_j=\theta+g^{-1}\!\left(r_{k}^{\mu}\right) & \text{if } r_{k}^{\mu}>0 \\
    \sum_j W_{kj} r^\mu_j\le\theta  & \text{if } r_{k}^{\mu}=0
\end{cases}
\end{align*}
The inequality constraints in this problem introduce additional degrees of freedom, making it harder to solve. The dense patterns case contains no inequalities as $r_{k}^{\mu}>0\ \forall\mu\forall k$, and hence is amenable to a fully analytic solution, \Cref{subsec:kkt}.}

\mysubsec{Numerical simulations details\label{subsec:numerics}}
\change{We have used networks with 128, 256, or 512 neurons in simulations, as indicated. Each point in the numerical curves of \Cref{fig:capacity} corresponds to a specific load value, $\alpha=P/N$, and was computed using the following procedure. We first sampled $P=\alpha N$ patterns independently (i.e., each activation value of each pattern is sampled i.i.d.), according to the pattern statistics used. Second, we created a network by either solving the numerical optimization problem for these patterns, \Cref{eq:optW}, or by using the analytic expression for dense patterns, \Cref{eq:denseW}, as detailed in the previous sections. Third, we measured the relevant statistic. This procedure was repeated $K=\max\{\left\lfloor1024/P\right\rfloor,1\}$ times (e.g., simulations with $P=4$ were repeated $256$ times, while simulations with $P\ge1024$ were done only once). }

\mysubsec{Measures of non-normal dynamics\label{subsec:non-normality}}
The asymmetry index is defined as $\left\|\mathbf{W}_\mathrm{asym}\right\|_\mathrm{F}/\left(\left\|\mathbf{W}_\mathrm{sym}\right\|_\mathrm{F}+\left\|\mathbf{W}_\mathrm{asym}\right\|_\mathrm{F}\right)$ using the symmetric and anti-symmetric parts of the weights $\mathbf{W}=\mathbf{W}_\mathrm{sym}+\mathbf{W}_\mathrm{asym}$.
The non-normality index is Henrici's deviation from normality index \cite{henrici1962bounds, trefethen2020spectra, stroud2023optimal} for the Jacobian $\mathbf{J}$, 
$\sqrt{\left\|\mathbf{J}\right\|_\mathrm{F}^2-\sum_i\left|\lambda_i\right|^2}/\left\|\mathbf{J}\right\|_\mathrm{F}$, where $\left\{\lambda_i\right\}_{i=1}^N$ are the Jacobian eigenvalues. 

\mysubsec{Optimal weights for dense patterns\label{subsec:kkt}} 
In the dense case, \Cref{eq:optW} can be solved in closed-form by denoting a Lagrangian:
$$
    {\cal L}=\frac{1}{2}\mathrm{Tr}\mathbf{W}^{T}\mathbf{W}+\mathrm{Tr}\left[\mathbf{\Gamma}^{T}\left(\mathbf{V}-\mathbf{W}\mathbf{R}\right)\right]+\boldsymbol{\gamma}^{T}\mathrm{diag}\left(\mathbf{W}\right)
$$
where $\mathbf{R},\mathbf{V}\in\mathbb{\mathbf{R}}^{N\times P}$ as defined in the main text, $\mathbf{\Gamma}\in\mathbb{\mathbf{R}}^{N\times P}$ are Lagrange multipliers enforcing the fixed-point conditions, and $\boldsymbol{\gamma}\in\mathbb{\mathbf{R}}^{N}$ are Lagrange multipliers enforcing lack of self-coupling in the connectivity. Then the optimal weights satisfy $0=\frac{\partial{\cal L}}{\partial \mathbf{W}}$, $0=\mathbf{V}-\mathbf{W}\mathbf{R}$, and $0=\mathrm{diag}\left(\mathbf{W}\right)$, which we solve for $\mathbf{\Gamma}$:
\begin{align*}
    \mathbf{W}	=& \mathbf{\Gamma} \mathbf{R}^{T}-\boldsymbol{\gamma}\circ \mathbf{I}\\
\mathbf{V}	=& \mathbf{W}\mathbf{R}=\mathbf{\Gamma} \mathbf{R}^{T}\mathbf{R}-\boldsymbol{\gamma}\circ \mathbf{R}\\
\mathbf{\Gamma}	=& \mathbf{V}\left(\mathbf{R}^{T}\mathbf{R}\right)^{-1}+\boldsymbol{\gamma}\circ \mathbf{R}\left(\mathbf{R}^{T}\mathbf{R}\right)^{-1}
\end{align*}
so that substituting $\mathbf{\Gamma}$ we have an expression for $\mathbf{W}$ in terms of the pseudo-inverse $\mathbf{R}^\dagger=\left(\mathbf{R}^{T}\mathbf{R}\right)^{-1}\mathbf{R}^\mathsf{T}$:
\begin{equation}\label{eq:optWzeroDiagonal}
    \mathbf{W}=\mathbf{V}\mathbf{R}^{\dagger}-\boldsymbol{\gamma}\circ\left(\mathbf{I_{N}}-\mathbf{R}\mathbf{R}^{\dagger}\right)
\end{equation}
and $\gamma_i$ is given by the equation $W_{ii}=0$:
$$
\gamma_{i}=\left[\mathbf{V}\mathbf{R}^{\dagger}\right]_{ii}\big/\left[\mathbf{I_{N}}-\mathbf{R}\mathbf{R}^{\dagger}\right]_{ii}
$$
Finally, without avoiding self-coupling in the connectivity $\mathbf{W}$, we have $\boldsymbol{\gamma}=0$ and recover \Cref{eq:denseW}.

\mysubsec{Stability transition in paired Gaussian matrices\label{subsec:lambda-bulk}}
As we previously showed \cite{cohen2025eigenvalue}, for a pair of rectangular matrices $\mathbf{X},\mathbf{Y}\in\mathbb{R}^{N\times P}$ for $\alpha=P/N<1$, whose corresponding entries are jointly Gaussian, i.e., any $\left(x,y\right)=\left(X_{i\mu},Y_{i\mu}\right)$ are i.i.d. 
$\left(x,y\right)\sim{\cal N}\left(0,\left(\begin{array}{cc}
\sigma_{x}^{2} & \tau\,\sigma_\mathrm{x}\,\sigma_\mathrm{y}\\
\tau\,\sigma_\mathrm{x}\,\sigma_\mathrm{y} & \sigma_\mathrm{y}^{2}
\end{array}\right)\right)$,
the support of the eigenvalue spectrum of $\mathbf{M}=\mathbf{X}\,\mathbf{Y}^\dagger$ for $N,P\to\infty$ is given by a circular law:
\begin{equation}\label{eq:GaussianPseudoinverseSupport}
\left|\lambda-\tau\,\frac{\sigma_\mathrm{x}}{\sigma_\mathrm{y}}\right|^2 \le \frac{\sigma_\mathrm{x}^2}{\sigma_\mathrm{y}^2}\,\left(1-\tau^2\right)\,\frac{\alpha}{1-\alpha}
\end{equation}
which is quite precise already for moderate $N$ (\Cref{fig:SM_GaussianPseudoinverse}). 

As a corollary, for $\mathbf{M}	= -c\,\mathbf{I}+\mathbf{X}\,\mathbf{Y}^{\dagger}$, the upper and lower bounds of $\mathrm{Re}\lambda$ are achieved at real numbers $\lambda_\pm$:
\begin{equation}\label{eq:GaussianPseudoinverseLambdaSupport}
    \lambda_{\pm} =-c+\frac{\sigma_\mathrm{x}}{\sigma_\mathrm{y}}\,\left(\tau\pm\sqrt{\frac{\alpha}{1-\alpha}}\,\sqrt{1-\tau^2}\right)
\end{equation}
yielding, for $c=1$, \Cref{eq:lambda_bulk} in the main text where $\lambda_+$ is called $\lambda_\mathrm{bulk}$.

Noting that $\alpha/\left(1-\alpha\right)$ is strictly monotonic for $\alpha\in\left(0,1\right)$, we can denote by $\alpha_\mathrm{S}$ the largest $\alpha$ where $\lambda_{+}<0$, 
$\alpha_\mathrm{S}=\max_{\lambda_{+}\left(\alpha\right)<0} \alpha$, and solve from the condition $\lambda_+<0$:
\begin{align*}
    \sqrt{\frac{\alpha}{1-\alpha}} &< \frac{c\,\sigma_\mathrm{y}^2-\tau\,\sigma_\mathrm{x}\,\sigma_\mathrm{y}}{\sigma_\mathrm{x}\,\sigma_\mathrm{y}\,\sqrt{1-\tau^2}}\\
    \alpha &<\frac{\left(c\,\sigma_\mathrm{y}^2-\tau\,\sigma_\mathrm{x}\,\sigma_\mathrm{y}\right)^2}{\sigma_\mathrm{x}^2\,\sigma_\mathrm{y}^2\,\left(1-\tau^2\right)+\left(c\,\sigma_\mathrm{y}^2-\tau\,\sigma_\mathrm{x}\,\sigma_\mathrm{y}\right)^2}
\end{align*}
so that when $c\sigma_{y}^{2}-\tau\,\sigma_{x}\,\sigma_{y}<0$ this inequality does not have a solution, and we can express $\alpha_\mathrm{S}$ compactly as
\begin{equation}\label{eq:GaussianPseudoinverseStability}
    \alpha_\mathrm{S}=\frac{\max\left(0,c\,\sigma_{y}^2-\tau\,\sigma_\mathrm{x}\,\sigma_\mathrm{y}\right)^2}{\sigma_\mathrm{x}^2\,\sigma_\mathrm{y}^2\,\left(1-\tau^2\right)+\left(c\,\sigma_\mathrm{y}^2-\tau\,\sigma_\mathrm{x}\,\sigma_\mathrm{y}\right)^2}
\end{equation}
so \Cref{eq:alphaSbulk} in the main text follows for $c=1$.

\mysubsec{An outlier related to the Jacobian average value\label{subsec:lambda-ave}}
To empirically find the outlier related to the Jacobian average value, we note that if all the eigenvectors were real, this outlier would have been associated with the eigenvector of J closest to the uniform vector $\boldsymbol{1}$:
\begin{equation}
v_{ave}	=\arg\max_{v\in \left\{ \mathrm{EV}(J):\|v\|=1\right\} }\boldsymbol{1}^{T}v
\end{equation}
Generalizing this idea to complex eigenvectors, we find the eigenvector $v$ such that the projection of the uniform vector $\boldsymbol{1}$ onto the span of $\mathrm{Re}\left(v\right), \mathrm{Im}\left(v\right)$ is of maximal norm. Denoting $V=\left[\mathrm{Re}\left(v\right), \mathrm{Im}\left(v\right)\right]\in\mathbb{R}^{N\times2}$ we have:
\begin{equation}
v_{ave}	=\arg\max_{v\in \left\{ \mathrm{EV}(J):\|v\|=1\right\} }\boldsymbol{1}^{T}VV^+\boldsymbol{1}
\end{equation}
where $V^+$ denotes the Moore-Penrose pseudo-inverse. Then $\lambda_\mathrm{ave}^\mathrm{emp}$ is the eigenvalue associated with $v_{ave}$.

\mysubsec{An outlier related to the memory pattern\label{subsec:lambda-mem}}
When for some constant $c$, $\mathbf{r^{\mu}}+c\mathbf{1}$ is an eigenvector of the Jacobian $J^{\mu}$ (\Cref{eq:FixedPointJacobian}) with eigenvalue $\lambda_{\mathrm{mem}}$:\looseness=-1
$$ \lambda_{\mathrm{mem}}\left(\mathbf{r^{\mu}}+c\mathbf{1}\right)	=-\mathbf{r^{\mu}}+\mathbf{g'\!\left(g^{-1}\!\left(r^\mu\right)\right)}\circ \mathbf{W}\mathbf{r^{\mu}}+c\lambda_{\mathrm{ave}}
$$
so denoting $\phi\left(x\right)=g'\!\left(g^{-1}\!\left(x\right)\right)\circ\left(g^{-1}\!\left(x\right)+\theta\right)$ we have
$$
\left(\lambda_{\mathrm{mem}}+1\right)\mathbf{r^{\mu}}+\lambda_{\mathrm{mem}}c\mathbf{1}	\\
=\mathbf{\boldsymbol{\phi}\left(r^\mu\right)}+c\lambda_{\mathrm{ave}}\mathbf{1}    
$$
a linear relation between $\mathbf{r^\mu}$ and $\mathbf{\boldsymbol{\phi}\left(r^\mu\right)}$, characterized by a Pearson correlation close to $1$, \Cref{eq:tauMem}, and in this case $\lambda_{\mathrm{mem}}+1$ is given by the linear regression slope, \Cref{eq:lambdaMem}.

\mysubsec{Stability dependence on the threshold\label{subsec:threshold}} 
In \Cref{eq:alphaSbulk}, $\alpha_\mathrm{S}^\mathrm{bulk}$ dependence on $\sigma$, $n$, and $\mathrm{CV}$ is hidden within the three (co)variance terms and is, in general, complex. Less so for $\theta$: $c_\mathrm{rr}$ do not include $\theta$ at all, and $c_\mathrm{rf}$ is independent of $\theta$ because
$c_\mathrm{rf} = \left\langle \delta r\, \delta g'\!\left(g^{-1}\!\left(r'\right)\right)\,g^{-1}\!\left(r\right)\right\rangle+\,\theta \left\langle \delta r\, \delta g'\!\left(g^{-1}\!\left(r'\right)\right)\right\rangle$, where the right term is $0$ due to the independence of $r,r'$. As $c_\mathrm{ff}$ depends on $\theta$, so does $\alpha_\mathrm{S}^\mathrm{bulk}$, but the condition $\alpha_\mathrm{S}^\mathrm{bulk}>0$ is independent of $\theta$ as it depends only on the numerator of \Cref{eq:alphaSbulk}, or $c_\mathrm{rr}>c_\mathrm{rf}$.
The zero crossing of $\lambda_\mathrm{ave}$ and $\lambda_\mathrm{mem}$ are expressed in \Cref{eq:theta_lambdaAve,eq:theta_lambdaMem}, respectively. The former is derived from \Cref{eq:lambdaAve}, and the latter from \Cref{eq:lambdaMem} by noting that $c_\mathrm{r\phi} = \left\langle \delta r\, \delta g'\!\left(g^{-1}\!\left(r\right)\right)\,g^{-1}\!\left(r\right)\right\rangle+\,\theta \left\langle \delta r\, \delta g'\!\left(g^{-1}\!\left(r\right)\right)\right\rangle$.

\begin{acknowledgments}
This work was supported by the Wellcome Trust (Investigator Award in Science 212262/Z/18/Z to M.L.), the Human Frontiers Science Programme (Research Grant RGP0044/2018 to M.L.), and the Blavatnik Cambridge Postdoctoral Fellowships (to U.C.). We thank Alessandro Treves, Yasser Roudi, Yashar Ahmadian, and Samuel Eckmann for useful discussions.
\end{acknowledgments}
\paragraph*{Author Contributions}{U.C.\ and M.L.\ designed research; U.C.\ performed research; U.C.\ and M.L.\ interpreted results; U.C.\ and M.L.\ wrote the paper.}
\paragraph*{The authors declare no competing interest.}
\paragraph*{Code availability} All code used to generate the included figures will be available in a persistent repository upon publication \cite{cohen2026github}.

\bibliography{biblio}

\setcounter{figure}{0} 
\renewcommand{\thefigure}{S\arabic{figure}} 

\onecolumngrid

\appendix

\section{Simulation results for sparse patterns}\label{subsec:sparse-patterns} 
\begin{figure}[H]
\centering
\includegraphics[width=\textwidth]{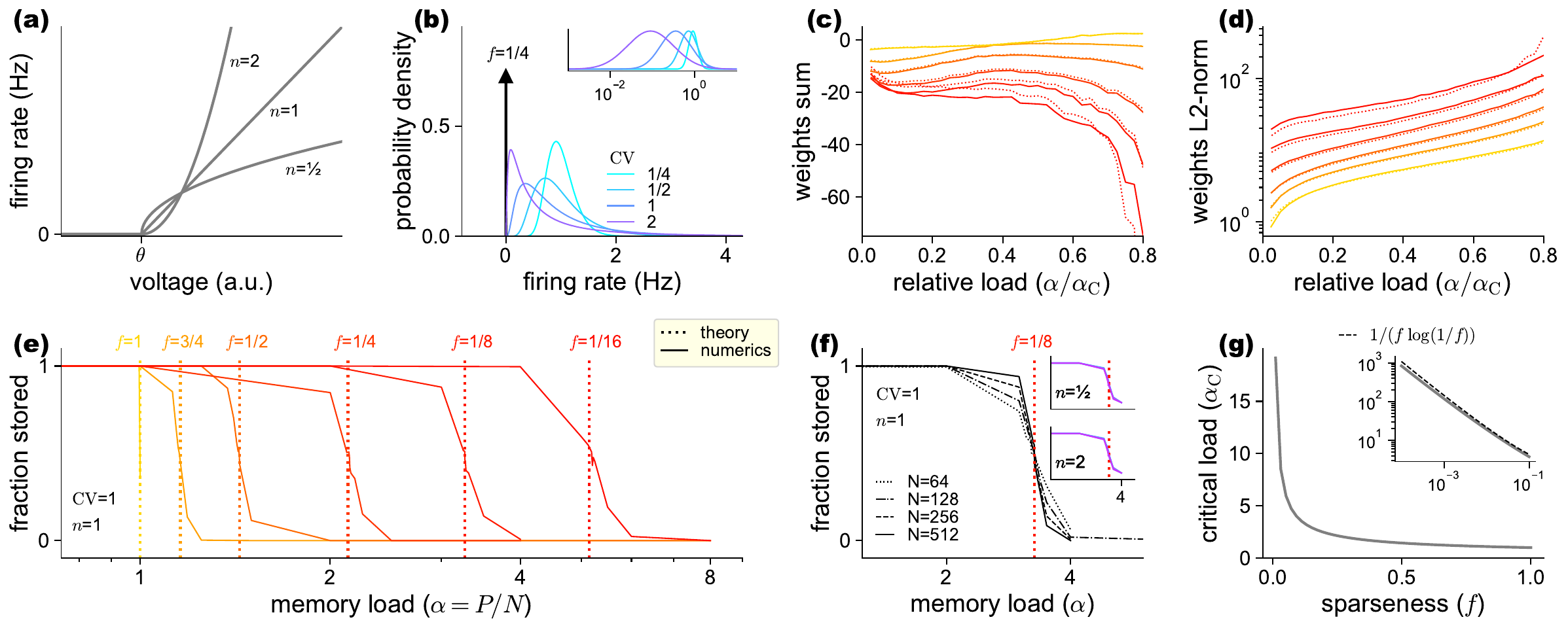}
\caption{\label{fig:extdata_fig_1} \textbf{Capacity phase-transition for sparse patterns.} (a) Rectified power activation with threshold $\theta$ and different exponents $n$\change{, for the noiseless case $\sigma=0$}. (b) Probability density of sparse, log-normal distributed patterns at different variation levels (CV, color coded). A fraction $1-f$ of the density is exactly at 0 (arrow). The inset shows normalized density on a log scale. (c,d) Weights matrix row sum (c) and L2-norm (d) at different sparseness levels (color coded; empirical - full line, theory - dotted line) and loads (x-axis). (e-f) The fraction of patterns that are correctly stored in simulations (full line) at different loads (x-axis), and different sparseness levels (e, color coded), or different values of $N$ (f, see legend), or different values of $n$ and $\textrm{CV}$ (f, insets). Theory's predictions on the critical load - dotted lines. (g) The relation between $f$ and $\alpha_\mathrm{C}$ in linear scale or log-log scale (inset). \change{As noted in the text, this relation is independent of other parameters, namely patterns variation $\mathrm{CV}$ and activation function details (exponent $n$, smoothness $\sigma$, and threshold $\theta$); see insets in (f), and in \Cref{fig:capacity}m-o.} \change{Noiseless activations, $\sigma=0$, were used throughout this figure.} \change{Numerical simulations optimized the weights $\mathbf{W}$ according to \Cref{eq:optW} at $N=256$, unless when specificed otherwise.}} 
\end{figure}

For sparse patterns with sparseness $f\in\left(0,1\right]$, a fraction $1-f$ of the pattern is exactly zero while the rest is log-normal distributed with a mean of $1$ and is parametrized by its coefficient of variation ($\mathrm{CV}$, \Cref{fig:extdata_fig_1}b). For these patterns, the relevant activation function has $\sigma=0$, and is parametrized by its exponent, $n$, and threshold, $\theta$ (\Cref{fig:extdata_fig_1}a). 
In this case, there is no closed-form expression for the optimal solution to \Cref{eq:optW}, but it can be found numerically using off-the-shelf tools \cite{Clarabel_2024}. The resulting weights matrix row sum and L2-norm match the prediction of the replica theory very well (\Cref{fig:extdata_fig_1}c-d). The maximal number of patterns which can be stored $\alpha_C$ depends only on $f$, not on other parameters  $n$, $\theta$, $\textrm{CV}$ or $N$ (\Cref{fig:extdata_fig_1}e-f) and scales asymptotically as $1/f\log\left(1/f\right)$, as in \cite{schonsberg2021efficiency} (\Cref{fig:extdata_fig_1}g).

\begin{figure}[H]
\centering
\includegraphics[width=\textwidth]{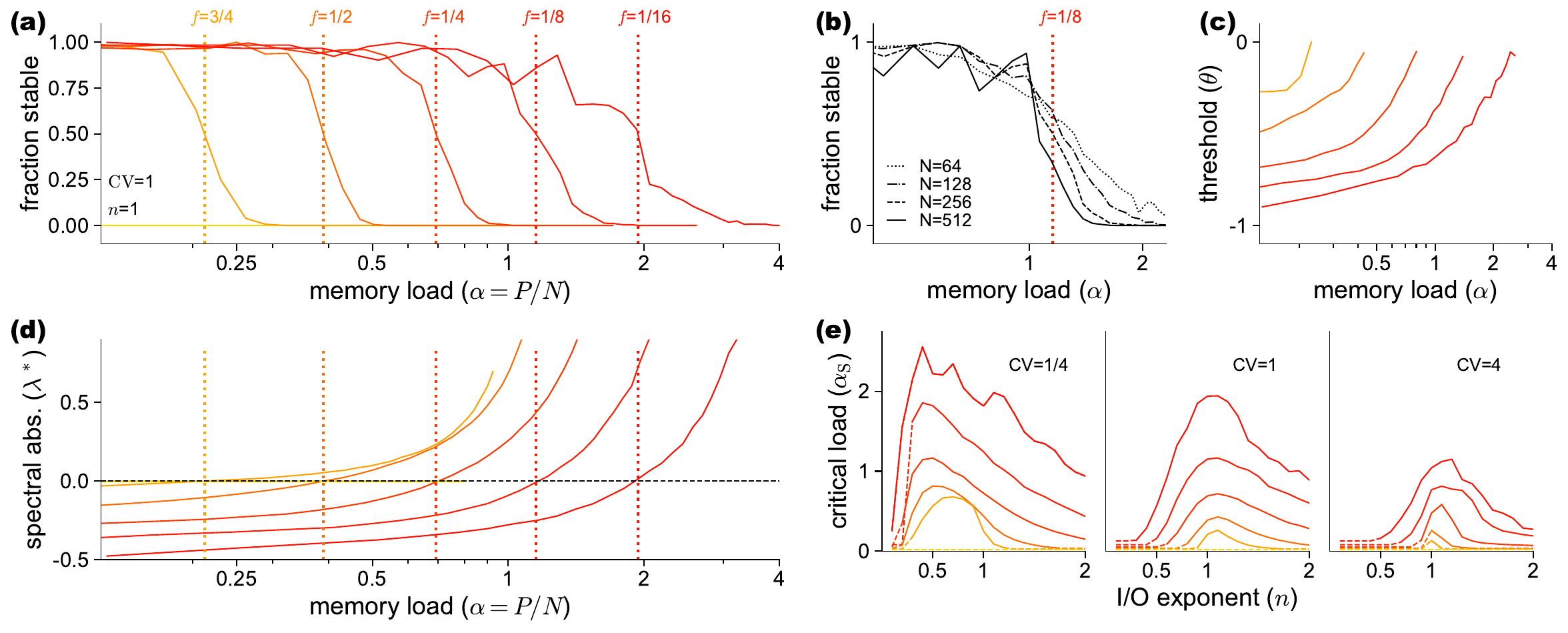}
\caption{\label{fig:extdata_fig_2} \textbf{Stability phase-transition for sparse patterns.} (a-b) The fraction of patterns which are stable for recall in simulations (full line) at different loads (x-axis), and different sparseness levels (a, color coded), or different values of $N$ (b, see legend). The empirically found critical load $\alpha_\mathrm{S}$ - dotted lines. (c-d) The optimal choice for threshold $\theta$ (c) and mean spectral abscissa $\lambda^*$ (the real part of the eigenvalue with the largest real part) (d) at different loads (x-axis), and different sparseness levels (color coded), for the same experiments as a. (e) The empirically found critical load for stability $\alpha_\mathrm{S}$ at different levels of variation (panels), different I/O exponents (x-axis), and levels of sparseness (color coded). Dashed lines connect load values below the minimal value identifiable by the experiments. \change{Noiseless activations, $\sigma=0$, were used throughout this figure.} \change{Numerical simulations optimized the weights $\mathbf{W}$ according to \Cref{eq:optW} at $N=256$, unless when specificed otherwise.}} 
\end{figure}

To quantify stability, we define the spectral abscissa as the median, over the stored patterns, of the eigenvalue with the largest real value. Optimizing the threshold $\theta$ for pattern stability (e.g., using a line search or finite differentiation with respect to the spectral abscissa), there is an evident stability phase transition at a critical load $\alpha_\mathrm{S}$ which increases with sparseness $f$ (\Cref{fig:extdata_fig_2}a, compare with \Cref{fig:extdata_fig_1}e), with the transition becoming steeper with $N$ (\Cref{fig:extdata_fig_2}b, compare with \Cref{fig:extdata_fig_1}f). For dense patterns, $f=1$, no memory pattern is stable, in agreement with the limit $\sigma\to0$ extrapolated from \Cref{fig:phase-diagram}a-b (left column).
The optimal threshold (empirically found) is always negative (\Cref{fig:extdata_fig_2}c), as well as the resulting weights sum at the optimal threshold (\Cref{fig:extdata_fig_1}c), and the spectral abscissa increases monotonically with the load (\Cref{fig:extdata_fig_2}d).

Unlike the critical load for storage (or `storage capacity') $\alpha_\mathrm{C}$, which depends only on $f$, the critical load for stability $\alpha_\mathrm{S}$ depends also on pattern statistics $\textrm{CV}$ and activation function exponent $n$ (at the optimal threshold $\theta$). Interestingly, at low levels of pattern variation $\textrm{CV}\le1$ the optimal exponent is sublinear $n^*<1$, while for high levels of pattern variation $\textrm{CV}\ge1$ the optimal exponent is linear, $n^*\approx1$ (\Cref{fig:extdata_fig_2}e). Thus, for binary patterns where $\textrm{CV}=0$, $n^*\approx0$ is optimal, as was the choice in [Hopfield (1982)] \cite{hopfield1982neural}. On the other hand, for high-variance patterns, we predict an optimal $n^*\approx 1$.

\section{Supplementary figures}\label{subsec:supp-figures}
\begin{figure}[H]
\centering
\includegraphics[width=\linewidth]{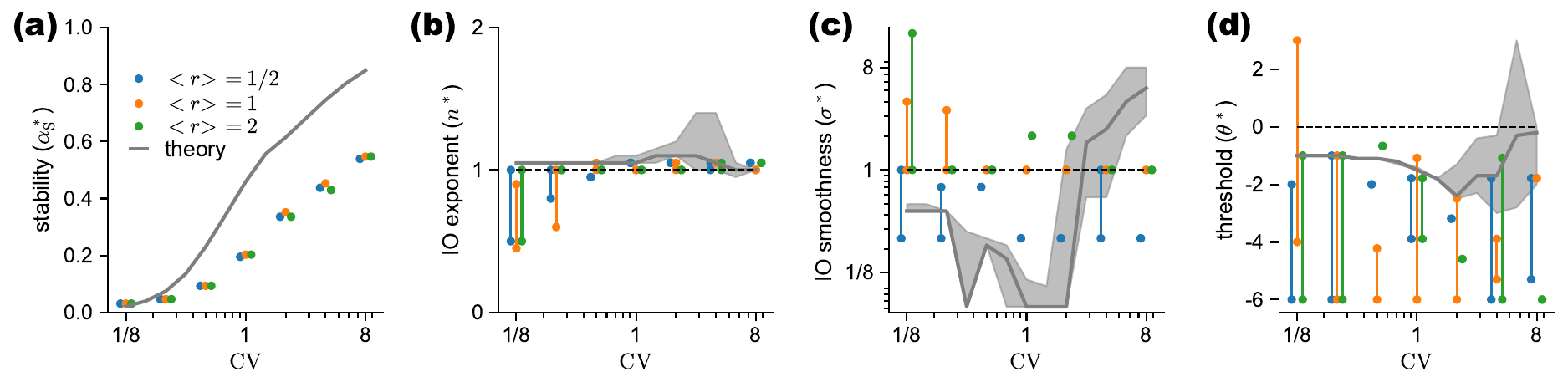}
\caption{\label{fig:sm_optimals}
\textbf{Optimal parameters.} (a) The critical load for stability using optimal parameters, at different pattern variation levels (x-axis), per theory (gray line) or simulations with different choices of $\left<r\right>$ (dots). (b-d) The optimal values (achieving $\alpha_\mathrm{S}^*$ in a) of I/O smoothness, $\sigma$ (b), I/O exponent, $n$ (c), and threshold, $\theta$ (d) at different pattern variation levels (x-axis), color-coded as in a. The range of values achieving 97.5\% of the optimum is indicated by a vertical line (simulations) and shading (theory). \change{Numerical simulations optimized the weights $\mathbf{W}$ according to \Cref{eq:denseW} at $N=256$.}
}
\end{figure}

\begin{figure}[H]
\centering
\includegraphics[width=\linewidth]{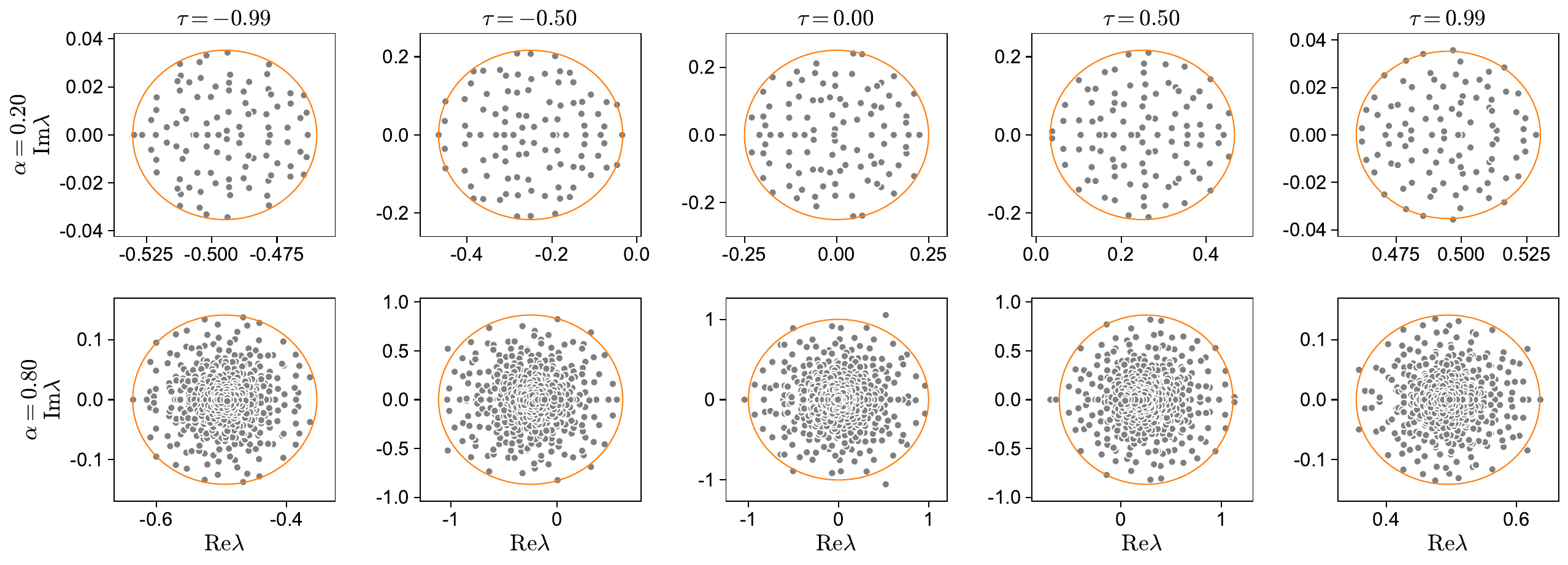}
\caption{\label{fig:SM_GaussianPseudoinverse}
\textbf{Eigenvalue spectrum of the Gaussian pseudo-inverse ensemble, $\mathbf{M}=\mathbf{X}\,\mathbf{Y}^\dagger$.} Eigenvalues (gray dots) and predicted support (orange contour, \Cref{eq:GaussianPseudoinverseSupport}) at different values of $\alpha$ (rows) and $\tau$ (columns). We used $N=500$, $\sigma_\mathrm{x}=1$, $\sigma_\mathrm{y}=2$.
Note the difference between panels in the range of the axes, as predicted by theory.}
\end{figure}

\begin{figure}[H]
\centering
\includegraphics[width=\linewidth]{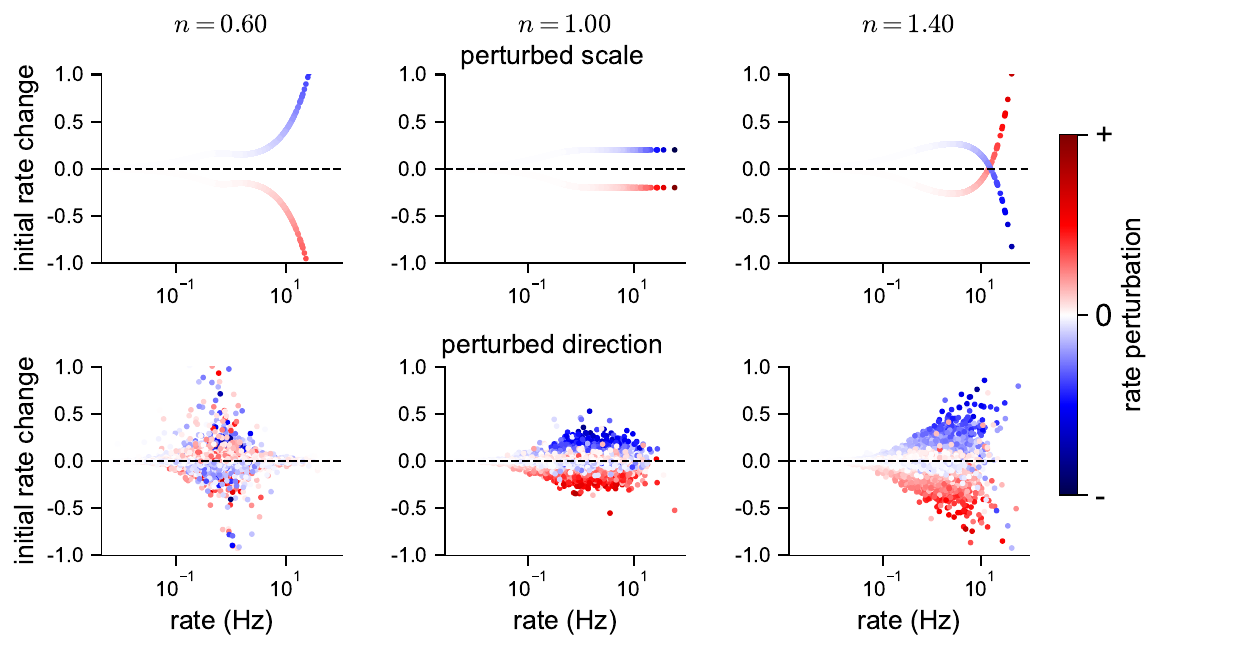}
\caption{\label{fig:sm_intuition}
\textbf{The effect of activation function nonlinearity on robustness to perturbations.} \change{Scatter plots show initial activity change after perturbation, $\dot{r}_i(t=0)$ (y-axis), against fixed-point pattern activation values before perturbation, $r^\mu_i$ (x-axis, log scale), for different values of the exponent of the activation function nonlinearity ($n$, columns) and perturbation type (rows). Each dot is one of $N=128$ neurons in one of $P=24$ stored patterns, with color showing direction and magnitude of perturbation, $r_i\!\left(t=0\right)-r^\mu_i$ (blue: decreased activity; white: no change; red: increased activity, see below for exact ranges). Note that (assuming the direction of initial change plotted here is representative of asymptotic behavior) stability requires blue dots to be above, and red dots to be below zero (dashed horizontal line).
Scale perturbations (top row) are defined as $\mathbf{r}\!\left(t=0\right)=\left(1\pm\varepsilon\right) \, \mathbf{r}^\mu$, and are presented with color ranges of $\pm6.54$, $\pm6.03$, and $\pm5.93$~Hz (left to right). Direction perturbations (bottom row) are defined as $\mathbf{r}\!\left(t=0\right)=g\!\left(\mathbf{v}^\mu + \varepsilon \, \delta\mathbf{v} \right) \, \frac{\left\|g(\mathbf{v}^\mu)\right\|}{\left\|g\!\left(\mathbf{v}^\mu+\varepsilon \, \delta\mathbf{v}\right)\right\|}$ for a vector of standard Gaussian noise $\delta\mathbf{v}$, and are presented with color ranges $\pm0.23$, $\pm0.41$, $\pm0.97$~Hz (left to right).
We used patterns with $\mathrm{CV}=2$, neurons with $\sigma=1$, $\theta=-2$, and a perturbation size of $\varepsilon=0.1$.}}
\end{figure}
\clearpage

\section{Replica theory for the classical storage capacity}\label{subsec:replica-theory}
We consider $P$ graded patterns $\mathbf{r^\mu}\in\mathbb{R}_+^N$ for $\mu=1\ldots P$, with sparseness level of $f\in\left(0,1\right]$, so that a fraction $1-f$ of all entries are exactly 0. We develop a replica theory for the ability to satisfy the $P$ non-linear equations $\mathbf{g}\left(\mathbf{W} \mathbf{r^{\mu}}\right)=\mathbf{r^{\mu}}$, defining $P$ fixed points for the dynamics of \Cref{eq:dynamics}, for any monotonic activation function $\ratefun\left(\cdot\right)$ with a threshold $\theta$ (cf.\ \Cref{eq:noisy_nonlinearity}). As the problem decouples for different rows, we denote the volume of solutions for a single row $\mathbf{w}=\mathbf{w^k}$ of the matrix $\mathbf{W}$, for any $k=1\ldots N$:
\begin{equation}\label{eq:solutions-volume}
{\cal V}=\left\{ \mathbf{w}: \left(r_{k}^{\mu}>0\ \cap\ \mathbf{w}^\mathsf{T} \mathbf{r^\mu}=\theta+g^{-1}\!\left(r_{k}^{\mu}\right)\right)\ \cup\ \left(r_{k}^{\mu}=0\ \cap\ \mathbf{w}^\mathsf{T} \mathbf{r^\mu}\le\theta\right)\right\} 
\end{equation}
This framing captures both the dense case where $f=1$ and the activation is strictly monotonic $g\left(\cdot\right):\mathbb{R}\to\mathbb{R}_+$, and the sparse case where we assume that the activation is rectified, i.e., $g\left(\cdot\right):\mathbb{R}_+\to\mathbb{R}_+$ is strictly monotonic with $g(0)=0$ and define $g(x)=0$ for any $x<0$. The notation $g^{-1}\!\left(\cdot\right)$ is defined only where $g\left(\cdot\right)$ is strictly monotonic. 

By characterizing the conditions where the volume of solutions $V$ vanishes, and correlation between different solutions peaks, we capture the unique minimal-norm solution corresponding to \Cref{eq:optW} in any finite $\alpha=P/N$. Intuitively, in this regime, only a single weight matrix solves the equations. As usual, it is sufficient to find $G^*$ such that $\left[V^n\right]=e^{nG^*}$ as invoking the replica identify $\left[\log V\right]=\lim_{n\to0}\frac{1}{n}\left(\left[V^{n}\right]-1\right)$ and L'Hôpital's rule implies $\left[\log V\right]=G^*$.

We start by writing the replicated volume in terms of Dirac $\delta$, Kronecker $\delta$, and the Heaviside step function $\Theta$:
$$
V^{n}	=\int d^{N\times n}w_{i}^{\alpha}\prod_{\alpha}^{n}\prod_{\mu}^{P}\left(\left(1-\delta_{r^{\mu}_k}\right)\delta\left(\theta+g^{-1}\!\left(r^{\mu}_k\right)-\sum_{i}^{N}w_{i}^{\alpha}r_{i}^{\mu}\right)+\delta_{r^{\mu}_k}\Theta\left(\theta-\sum_{i}^{N}w_{i}^{\alpha}r_{i}^{\mu}\right)\right)
$$
which we seek to average over the i.i.d. sampling of the patterns $r^\mu_i$, noting it can be done independently assuming there are no self-couplings $W_{kk}=w_k=0$ so that terms with $r_{i}^{\mu}$ and $r_{k}^{\mu}$ are independent. We denote:
\begin{align*}
    I_{1} &= \left[\left[\prod_{\alpha}^{n}\delta\left(\theta+g^{-1}\!\left(r^{\mu}_k\right)-\sum_{i}^{N}w_{i}^{\alpha}r_{i}^{\mu}\right)\right]_{r^{\mu}}\right]_{r^{\mu}_k>0}\\
I_{2}	&=\left[\prod_{\alpha}^{n}\Theta\left(\theta-\sum_{i}^{N}w_{i}^{\alpha}r_{i}^{\mu}\right)\right]_{r^{\mu}}
\end{align*}
and use the Taylor expansion of $\left[ e^{x}\right]$  around $\left[x\right]$, $\left[ e^{x}\right] 	\approx e^{\left[ x\right] +\frac{1}{2}\left[ \left(\delta x\right)^{2}\right] }$, denoting the pattern statistics  $x_{1}=\left[r_{i}^{\mu}\right]$, $x_{2}=\left[\left(\delta r_{i}^{\mu}\right)^{2}\right]$, $y_{1}=\left[g^{-1}\!\left(r^\mu_k\right)|r^\mu_k>0\right]$, $y_{2}=\left[\left(\delta g^{-1}\!\left(r^\mu_k\right)\right)^{2}|r^\mu_k>0\right]$, and using the independence $\left[\delta r_{i}^{\mu}\delta r_{j}^{\mu}\right]=\delta_{ij}x_{2}$:
\begin{align*}
I_{1}	&=\int d^{n}\hat{s}^{\alpha}e^{i\sum_{\alpha}^{n}\hat{s}^{\alpha}\left(\theta+y_{1}-x_{1} m^{\alpha}\right)-\frac{1}{2}\sum_{\alpha\beta}^{n}\hat{s}^{\alpha}\hat{s}^{\beta}\left(x_{2} Q_{\alpha\beta}+y_{2}\right)}\\
I_{2}	&=\int_{-\infty}^{\theta}d^{n}k_{\alpha}\int d^{n}\hat{s}^{\alpha}e^{i\sum_{\alpha}^{n}\hat{s}^{\alpha}\left(x_{1} m^{\alpha}-k_{\alpha}\right)-\frac{1}{2}\sum_{\alpha\beta}^{n}\hat{s}^{\alpha}\hat{s}^{\beta}\left(x_{2} Q_{\alpha\beta}\right)}
\end{align*}
denoting new order parameters for the weight correlations,
$m^{\alpha}	=\sum_{i}^{N}w_{i}^{\alpha}$ and $Q_{\alpha\beta}	=\sum_{i}^{N}w_{i}^{\alpha}w_{i}^{\beta}$ (note the different scaling compared to \cite{schonsberg2021efficiency}). After $N$ decoupled $n$-dimensions Gaussian integrals on $w_{i}^{\alpha}$ we have:
\begin{align*}
    \left[V^{n}\right]_r	&=\int d^{n}m_{\alpha}\int\frac{d^{n}\hat{m}_{\alpha}}{2\pi}\int d^{n\times n}Q_{\alpha\beta}\int\frac{d^{n\times n}\hat{Q}_{\alpha\beta}}{2\pi}e^{NG}\\
G	&= \frac{1}{N}\sum_{\alpha}^{n}i\hat{m}_{\alpha}m_{\alpha}+\frac{1}{2N}\sum_{\alpha\beta}^{n}\left(2i\hat{Q}_{\alpha\beta}\right)Q_{\alpha\beta}-\frac{1}{2}\log\det\left(2i\hat{Q}\right)
	+\frac{1}{2}\sum_{\alpha\beta}^{n}i\hat{m}_{\alpha}i\hat{m}_{\beta}\left(2i\hat{Q}\right)_{\alpha\beta}^{-1}+f\alpha\log I_{1}+\left(1-f\right)\alpha\log I_{2}
\end{align*}

Assuming the replica symmetry ansatz $m_{\alpha}	=m$, $i\hat{m}_{\alpha}	=\hat{m}$, $Q_{\alpha\beta}	=q+\left(q_{0}-q\right)\delta_{\alpha\beta}$ and $2i\hat{Q}_{\alpha\beta}	=\hat{q}+\left(\hat{q}_{0}-\hat{q}\right)\delta_{\alpha\beta}$: 
\begin{align*}
     \left[V^{n}\right]_r	&=\int dm\int d\hat{m}\int dq_{0}\int dq\int d\hat{q}_{0}\int d\hat{q}e^{NnG}\\
G	&=\frac{1}{N}\hat{m}m+\frac{1}{2N}\left(q_{0}\hat{q}_{0}-q\hat{q}\right)-\frac{1}{2}\log\left(\hat{q}_{0}-\hat{q}\right)-\frac{1}{2}\frac{\hat{q}}{\hat{q}_{0}-\hat{q}}+\frac{1}{2}\frac{\hat{m}^{2}}{\hat{q}_{0}-\hat{q}}+\frac{f\alpha}{n}\log I_{1}+\frac{\left(1-f\right)\alpha}{n}\log I_{2}
\end{align*}

Now, using Hubbard-Stratonovich $e^{-y^{2}/2}=\int Dte^{ity}$ for $Dt=\frac{dt}{\sqrt{2\pi}}e^{-t^{2}/2}$ to get rid of the squares and decouple the n replicas, $I_1, I_2$ become:
\begin{align*}
    I_{1}&=\int d^{n}\hat{s}^{\alpha}e^{i\sum_{\alpha}^{n}\hat{s}^{\alpha}\left(\theta+y_{1}-x_{1}m\right)-\frac{1}{2}x_{2}\left(q_{0}-q\right)\sum_{\alpha}^{n}\left(\hat{s}^{\alpha}\right)^{2}-\frac{1}{2}\left(\sqrt{x_{2}q+y_{2}}\sum_{\alpha}^{n}\hat{s}^{\alpha}\right)^{2}}\\
    &=\int Dt\left(e^{-\frac{1}{2}\frac{\left(\theta+y_{1}-x_{1}m+t\sqrt{x_{2}q+y_{2}}\right)^{2}}{x_{2}\left(q_{0}-q\right)}-\frac{1}{2}\log\left(x_{2}\left(q_{0}-q\right)\right)}\right)^{n}\\
    I_{2}&=\int_{-\infty}^{\theta}d^{n}k_{\alpha}\int d^{n}\hat{s}^{\alpha}e^{i\sum_{\alpha}^{n}\hat{s}^{\alpha}\left(x_{1}m^{\alpha}-k_{\alpha}\right)-\frac{1}{2}x_{2}\left(q_{0}-q\right)\sum_{\alpha}^{n}\left(\hat{s}^{\alpha}\right)^{2}-\frac{1}{2}\left(\sqrt{x_{2}q}\sum_{\alpha}^{n}\hat{s}^{\alpha}\right)^{2}}\\
    &=\int Dt\left(\int_{-\infty}^{\theta}dke^{-\frac{1}{2}\frac{\left(x_{1}m-k+t\sqrt{x_{2}q}\right)^{2}}{x_{2}\left(q_{0}-q\right)}-\frac{1}{2}\log\left(x_{2}\left(q_{0}-q\right)\right)}\right)^{n}
\end{align*}
and using the replica trick $\log\int Dt\left(Z\left(t\right)\right)^{n}=n\int Dt\log Z\left(t\right)$ for $n\to0$ we have:
\begin{align*}
    \log I_{1}	&=-\frac{n}{2}\left(\frac{\left(\theta+y_{1}-x_{1}m\right)^{2}}{x_{2}\left(q_{0}-q\right)}+\frac{x_{2}q+y_{2}}{x_{2}\left(q_{0}-q\right)}+\log\left(x_{2}\left(q_{0}-q\right)\right)\right)\\
\log I_{2}	&=n\int Dt\log\int_{-\infty}^{\theta}dke^{-\frac{1}{2}\frac{\left(x_{1}m-k+t\sqrt{x_{2}q}\right)^{2}}{x_{2}\left(q_{0}-q\right)}-\frac{1}{2}\log x_{2}\left(q_{0}-q\right)}
\end{align*}

Taking the limit $N\to\infty$ we consider the saddle-point equations $0=\frac{\partial G}{\partial\hat{m}}=\frac{\partial G}{\partial\hat{q}}=\frac{\partial G}{\partial\hat{q}_{0}}$ which are solved for:
$$
    G^*=\frac{1}{2}-\frac{1}{2}\frac{m^{2}/N}{q_{0}-q}+\frac{1}{2}\log\left(q_{0}-q\right)+\frac{1}{2}\frac{q}{q_{0}-q}+\frac{\alpha f}{n}\log{I_1}+\frac{\alpha \left(1-f\right)}{n}\log{I_2}
$$

As we aim to recover the minimal norm weights, we are interested in the limit where the volume vanishes and only a single solution remains, which is captured by the limit $0=\lim_{q_0\to q}\left(q_0\to q\right)G^*$, defining $\theta_{0}=\frac{\theta-x_{1}m}{\sqrt{x_{2}q}}$ for short:
\begin{align*}
    \lim_{q_{0}\to q}\left(q_{0}-q\right)\frac{1}{n}\log I_{1} &=-\frac{1}{2}\left(\left(\frac{y_{1}}{\sqrt{x_{2}}}+\frac{\theta-x_{1}m}{\sqrt{x_{2}}}\right)^{2}+q+\frac{y_{2}}{x_{2}}\right)\\
    \lim_{q_{0}\to q}\left(q_{0}-q\right)\frac{1}{n}\log I_{2} &=-\frac{\alpha q}{2}\int_{\frac{\theta-x_{1}m}{\sqrt{x_{2}q}}}^{\infty}Dt\left(t-\frac{\theta-x_{1}m}{\sqrt{x_{2}q}}\right)^{2}\\
    \lim_{q_{0}\to q}\left(q_{0}-q\right)G^* &= \frac{q-m^{2}/N}{2}-f\frac{\alpha q}{2}\left(\theta_{0}+\frac{y_{1}}{\sqrt{x_{2}q}}\right)^{2}-f\frac{\alpha q}{2}\left(1+\frac{y_{2}}{x_{2}q}\right)-\left(1-f\right)\frac{\alpha q}{2}\int_{\theta_{0}}^{\infty}Dt\left(t-\theta_{0}\right)^{2}
\end{align*}

Noting that $m^2/N\ll q$ as $m\sim O(1)$ and substituting $0=\lim_{q_0\to q}\left(q_0\to q\right)G^*$ we have one equation relating the unknown quantities $q,\theta_0$ through the known quantities $\alpha,x_1,x_2,y_1,y_2$. Assuming the solution is achieved at a finite value of $q$, it would satisfy the saddle-point condition $0=\frac{\partial G}{\partial q}$, yielding a second equation. The resulting self-consistent equations become, solving the integral and denoting $\varphi\left(x\right)=e^{-x^{2}/2}/\sqrt{2\pi}$ and $H\left(x\right)=\int_x^\infty \varphi\left(x\right)$:
\begin{align*}
\alpha^{-1}	&=f\left(\theta_{0}+\frac{y_{1}}{\sqrt{x_2 q}}\right)^{2}+f\left(\frac{y_{2}}{x_2 q}+1\right)+\left(1-f\right)\left(\left(\theta_{0}^{2}+1\right)H\left(\theta_{0}\right)-\theta_{0}\varphi\left(\theta_{0}\right)\right)\\
0	&=f\left(\theta_{0}+\frac{y_{1}}{\sqrt{x_2 q}}\right)+\left(1-f\right)\left(\theta_{0}H\left(\theta_{0}\right)-\varphi\left(\theta_{0}\right)\right)\\
\theta	&=mx_{1}+\theta_{0}\sqrt{x_2 q}
\end{align*}

Those equations are considerably simplified for $f=1$, yielding Eq. \ref{eq:meanW}-\ref{eq:meanSqrW} for the moments of $\mathbf{W}$.
The equations for the critical storage capacity $\alpha_C$ are given by noting that the weights norm diverges in this case, $q\to\infty$:
\begin{align*}
\alpha_\mathrm{C}^{-1}	&=f\theta_{0}^{2}+f+\left(1-f\right)\left(\left(\theta_{0}^{2}+1\right)H\left(\theta_{0}\right)-\theta_{0}\varphi\left(\theta_{0}\right)\right)\\
0	&=f\theta_{0}+\left(1-f\right)\left(\theta_{0}H\left(\theta_{0}\right)-\varphi\left(\theta_{0}\right)\right)
\end{align*}
which can be solved numerically, first for $\theta_0$, then for $\alpha_C$. For the dense case, $f=1$, we have $\theta_0=0$, and $\alpha_\mathrm{C}=1$.

\end{document}